\documentclass[sigconf, nonacm]{acmart}

\newcommand\vldbdoi{XX.XX/XXX.XX}
\newcommand\vldbpages{XXX-XXX}
\newcommand\vldbvolume{14}
\newcommand\vldbissue{1}
\newcommand\vldbyear{2020}
\newcommand\vldbauthors{\authors}
\newcommand\vldbtitle{\shorttitle} 
\newcommand\vldbavailabilityurl{URL_TO_YOUR_ARTIFACTS}
\newcommand\vldbpagestyle{plain} 

\usepackage{amsmath}
\usepackage{algorithm}
\usepackage{booktabs}
\usepackage{listings}
\usepackage{algpseudocode}
\usepackage{graphicx}
\usepackage{diagbox}
\usepackage{multirow}
\usepackage{subcaption}
\usepackage{enumitem}
\usepackage{tikz}
\usetikzlibrary{arrows.meta, positioning}

\begin{document}
\title{SOLAR: Scalable Distributed Spatial Joins through Learning-based Optimization}

 \author{Yongyi Liu}
 \affiliation{%
   \institution{University of California, Riverside}
   \city{Riverside}
   \state{California}
   \country{}
 }
 \email{yliu786@ucr.edu}

 \author{Ahmed R. Mahmood}

\affiliation{%
   \institution{Google LLC.}
   \city{Mountain View}
   \state{California}
   \country{}
}
 \email{amahmoo@google.com}

 \author{Amr Magdy}
 \affiliation{%
   \institution{University of California, Riverside}
   \city{Riverside}
   \state{California}
   \country{}
 }
 \email{amr@cs.ucr.edu}

 \author{Minyao Zhu}
 \affiliation{%
   \institution{Google LLC.}
   \city{Mountain View}
   \state{California}
   \country{}
 }
 \email{minyaozhu@gmail.com}    

\begin{abstract}
The proliferation of location-based services has led to massive spatial data generation. Spatial join is a crucial database operation that identifies pairs of objects from two spatial datasets based on spatial relationships. Due to the intensive computational demands, spatial joins are often executed in a distributed manner across clusters. However, current systems fail to recognize similarities in the partitioning of spatial data, leading to redundant computations and increased overhead. Recently, incorporating machine learning optimizations into database operations has enhanced efficiency in traditional joins by predicting optimal strategies. However, applying these optimizations to spatial joins poses challenges due to the complex nature of spatial relationships and the variability of spatial data. This paper introduces \textit{SOLAR}, scalable distributed spatial joins through learning-based optimization. \textit{SOLAR} operates through offline and online phases. In the offline phase, it learns \textit{balanced} spatial partitioning based on the similarities between datasets in query workloads seen so far. In the online phase, when a new join query is received, \textit{SOLAR} evaluates the similarity between the datasets in the new query and the already-seen workloads using the trained learning model. Then, it decides to either reuse an existing partitioner, avoiding unnecessary computational overhead, or partition from scratch. Our extensive experimental evaluation on real-world datasets demonstrates that \textit{SOLAR} achieves up to 3.6X speedup in overall join runtime and 2.71X speedup in partitioning time compared to state-of-the-art systems.


\end{abstract}

\maketitle

\pagestyle{\vldbpagestyle}
\begingroup\small\noindent\raggedright\textbf{PVLDB Reference Format:}\\
\vldbauthors. \vldbtitle. PVLDB, \vldbvolume(\vldbissue): \vldbpages, \vldbyear.\\
\href{https://doi.org/\vldbdoi}{doi:\vldbdoi}
\endgroup
\begingroup
\renewcommand\thefootnote{}\footnote{\noindent
This work is licensed under the Creative Commons BY-NC-ND 4.0 International License. Visit \url{https://creativecommons.org/licenses/by-nc-nd/4.0/} to view a copy of this license. For any use beyond those covered by this license, obtain permission by emailing \href{mailto:info@vldb.org}{info@vldb.org}. Copyright is held by the owner/author(s). Publication rights licensed to the VLDB Endowment. \\
\raggedright Proceedings of the VLDB Endowment, Vol. \vldbvolume, No. \vldbissue\ %
ISSN 2150-8097. \\
\href{https://doi.org/\vldbdoi}{doi:\vldbdoi} \\
}\addtocounter{footnote}{-1}\endgroup

\ifdefempty{\vldbavailabilityurl}{}{
\vspace{.3cm}
\begingroup\small\noindent\raggedright\textbf{PVLDB Artifact Availability:}\\
The source code, data, and/or other artifacts have been made available at \url{https://github.com/Yongyi-Liu/SOLAR}.
\endgroup
}

\section{Introduction}
\label{introduction}
A spatial join, specifically a spatial theta join, is a database operation that identifies pairs of objects from two spatial datasets that satisfy a spatial predicate. This operation is extensively applied in various fields including urban planning~\cite{wu2023spatial}, i.e., overlaying demographic data over political district polygons for political districting, public safety~\cite{wier2009area}, i.e., identifying vehicle-pedestrian injury in each census tract by joining the collision data with census tract, and medical imaging~\cite{wang2012accelerating}, i.e., detecting proximate cells.
Given the vast amount of spatial data, spatial joins are increasingly processed in distributed computing environments to efficiently manage large-scale spatial data.

\textbf{Motivation:} In practice, enterprise customers tend to issue similar queries repeatedly~\cite{teevan2007information, huang2024sibyl,schmidt2024predicate}. Huang et al.~\cite{huang2024sibyl} point out that Microsoft workloads are highly recurrent. Schmidt et al.~\cite{schmidt2024predicate} further propose that in Amazon Redshift data warehouse, customer workloads exhibit highly similar query patterns, i.e., users and systems frequently issue the same queries. Similarly, in spatial data systems~\cite{eldawy2015spatialhadoop, aji2013hadoop, yu2019spatial, eldawy2021beast, xie2016simba, tang2016locationspark, tang2020locationspark}, users frequently perform spatial joins involving datasets with similar spatial characteristics.

A critical step in executing distributed spatial joins is constructing a balanced partitioner, a data structure used to distribute spatial data fairly across worker nodes. However, this partitioning process incurs substantial computational overhead~\cite{yang2020efficient}. For instance, partitioning just two terabytes of spatial data can take up to five hours~\cite{aly2015aqwa}. Due to the expensive nature of constructing a partitioner and the repeating query patterns in industry workloads, once an effective spatial partitioner is established, there is no need to recompute it for each occurrence of the same or similar query. However, existing systems~\cite{eldawy2015spatialhadoop,aji2013hadoop,yu2019spatial,eldawy2021beast,xie2016simba,tang2016locationspark,tang2020locationspark} recalculate the partitioner for these frequently executed joins, which results in repeating computations for similar queries. 
In the meantime, spatial joins often share inherent similarities. For example, the process of joining datasets of parks with restaurants often requires a similar partitioning to those used when joining datasets of parks with cafes, due to spatial correlations between these entities. Additionally, when working with an updated version of a dataset, such as points of interest, new or deleted records may be present. However, the overall distribution is likely to remain unchanged. By recognizing and leveraging these patterns, it is possible to apply a previously determined partitioner to similar joins. In this way, the system can bypass rescanning the input data to gather distribution statistics and skip building a partitioner on-the-fly (repartition). 

\begin{figure*}
  \centering
  \includegraphics[width=\linewidth]{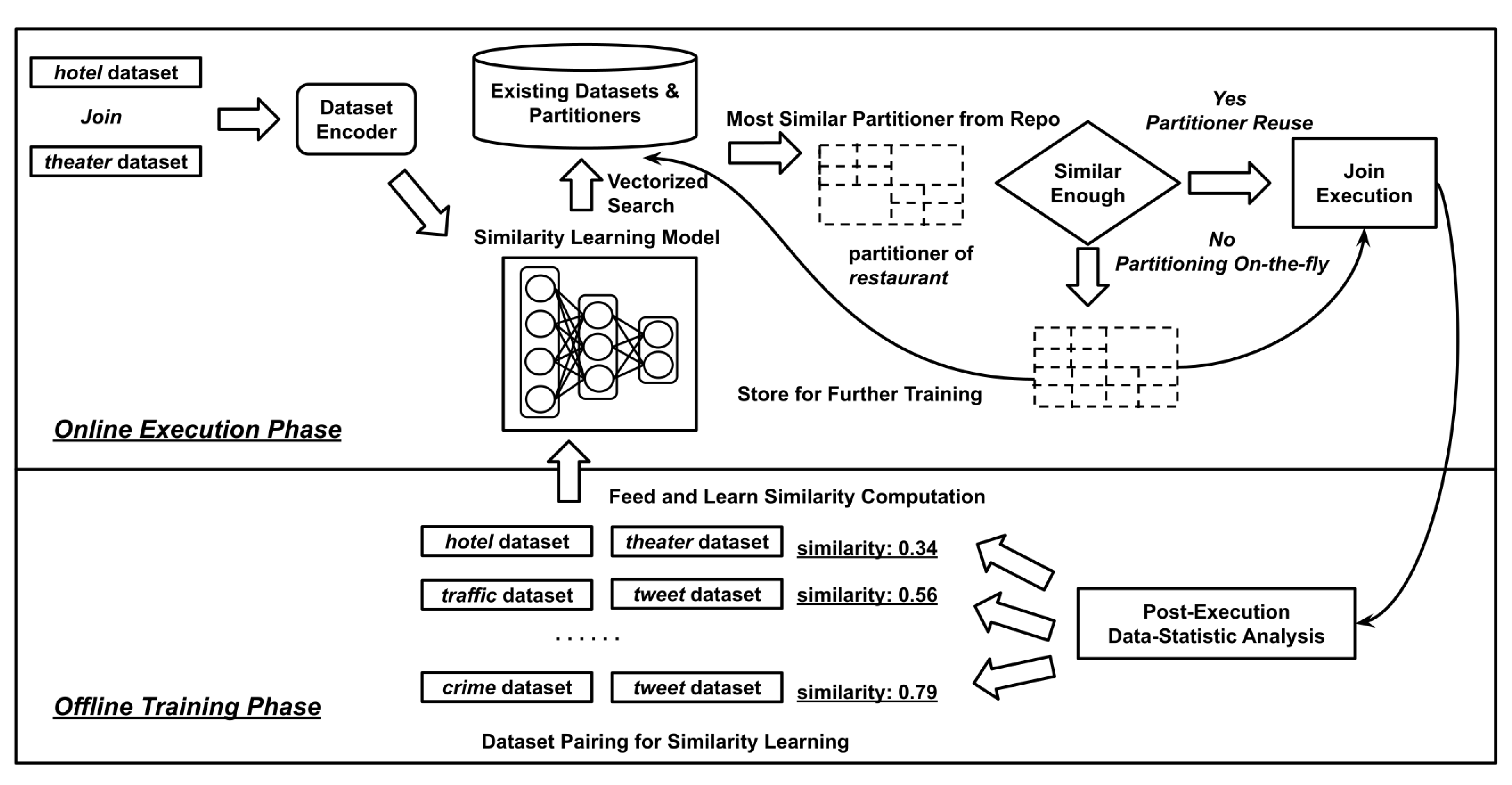}
  \vspace{-7mm}
  \caption{Example of \textit{SOLAR} Reusing Existing Partitioners in Executinng Spatial Join Queries}
  \label{fig:runningexample}
  \vspace{-4mm}
\end{figure*}

\textbf{Challenges:} 
Optimizing future spatial join workloads by leveraging historical spatial join workloads introduces four challenges.
(1) Despite the frequent occurrence of similar queries over identical or spatially similar datasets, there is no standardized metric to quantitatively evaluate the similarity between different spatial joins and their underlying spatial join inputs, i.e., datasets. Identifying accurate similarity metrics between inherently skewed spatial datasets and expensive spatial joins is challenging.
(2) Being able to learn the similarity between spatial joins and datasets calls for efficient embeddings of join queries and their corresponding datasets. This is very challenging as these embeddings must be compact for efficient storage, yet comprehensive enough to encapsulate the essential features of the queries and their corresponding inputs. 
(3) The utilization of historical spatial joins and their underlying spatial properties of join inputs to enhance the processing of new spatial joins adds another layer of complexity.  The challenge here is that the system needs to accurately and efficiently match a new spatial join query to a pre-existing spatially similar join or dataset in its history whenever possible.
(4) Making an accurate decision for an incoming spatial join that deviates from historical patterns—whether to adapt the partitioner from a historical spatial join workload or execute the new query without using historical information (i.e., repartition spatial join inputs)—is a very challenging problem. The reason is that making a wrong decision about reusing historical information could lead to poor spatial join performance, and avoiding partition reuse when needed will also incur unneeded spatial partitioning overhead.

\textbf{Existing work:} 
Recently, learning-based approaches have been widely used to optimize various operations in database systems. In non-spatial joins, reinforcement learning has been used to determine the optimal join orders of multiple tables~\cite{krishnan2018learning,marcus2018deep,yu2020reinforcement,yan2023join,chen2022efficient}. In spatial joins, reinforcement learning has been used to find an optimal partitioner~\cite{hori2023learned}, and traditional classification and regression models have been used to build spatial join optimizers~\cite{vu2020using,vu2024learning}, which can determine the best spatial join algorithm or partitioning method for a given query based on data distribution. However, none of these works attempts to reuse existing partitioners for future workloads of spatial join queries.

\textbf{Contribution:} To address the highlighted challenges and the limitations of existing work, we introduce \textit{SOLAR}, \underline{S}calable Distributed Spatial J\underline{o}ins through \underline{L}e\underline{ar}ning-based Optimization, that operates in both offline and online phases. 
In the offline phase, \textit{SOLAR} trains a Siamese Neural Network~\cite{bromley1993signature} to capture the overall spatial properties of datasets using the features extracted from dataset's metadata. This training shifts the model from depending on detailed data histograms to employing abstract dataset embeddings for similarity assessment. During the online phase, when a new join query is submitted, \textit{SOLAR} encodes the datasets involved in the query. It evaluates these datasets against those in the repository using the trained neural network to evaluate pairwise similarities using fast vector-based comparisons to retrieve the most similar dataset and its associated partitioner. This partitioner is reused to efficiently partition the new join's dataset.

Figure~\ref{fig:runningexample} illustrates how \textit{SOLAR} reuses existing partitioners. Consider a scenario where \textit{SOLAR} executes a join between the \textit{hotel} and \textit{theater} datasets. First, both datasets are embedded. Next, \textit{SOLAR}'s learned model performs a vectorized computation to locate the repository dataset most similar to either input. If the identified dataset is sufficiently similar, its existing partitioner is retrieved and reused; otherwise, a new partitioner is constructed on the fly (repartition) and stored in the repository. In this example, \textit{restaurant} is identified as sufficiently similar to \textit{hotel}, so \textit{SOLAR} reuses the \textit{restaurant} partitioner for the \textit{hotel}–\textit{theater} join. After the join is completed, data statistics (i.e., data histograms) are created for the input datasets to capture their underlying distributions. These histograms allow us to compute the ground-truth similarity for any pair of datasets in the repository. Such ground-truth values are then used to train \textit{SOLAR}'s similarity-learning model: the model sees only a dataset's metadata-based embedding and is trained to approximate the similarity measured by the histograms. This iterative process ensures that, over time, \textit{SOLAR} refines its ability to identify and reuse partitioners accurately based on metadata alone. In our experiments, whenever the decision maker opts to reuse a partitioner, \textit{SOLAR} consistently outperforms all baseline algorithms, achieving up to a 3.6X speedup in overall join runtime and a 2.71X speedup in partitioning time.

The main contributions of this paper are summarized as follows:
\begin{itemize}
    \item We propose \textit{SOLAR}, scalable distributed spatial joins through learning-based optimization that improves the performance of distributed spatial joins by reusing previously computed partitioners.
    \item We propose a method to compute the similarity between datasets based on data histograms. 
    \item We propose an embedding method for spatial datasets based on their metadata, i.e., polygon covering. 
    \item We propose a learning strategy to learn dataset similarities directly from dataset metadata, bypassing the need to compute histograms for each query.
    \item We perform experiments with various data sources, demonstrating that \textit{SOLAR} significantly improves the efficiency of distributed spatial joins compared to all spatial join partitioning and join algorithms of state-of-the-art system.
\end{itemize}

The rest of this paper is structured as follows:
Section~\ref{relatedwork} outlines the related work.
Section~\ref{problem} presents the preliminaries and problem definition.  Section~\ref{partitioner} details the selection of the partitioner in our system. Section~\ref{sim_metric} describes the similarity evaluation metrics between datasets. Sections~\ref{offline} and Section~\ref{online} discuss the offline and online phases of \textit{SOLAR}, respectively. Section~\ref{experiment} presents experimental evaluation, and Section~\ref{conclusion} concludes the paper and outlines future work.


\section{Related Work}
\label{relatedwork}

\textbf{Distributed Spatial Joins:}  Distributed spatial join typically follows a two-step process~\cite{yang2020efficient,yang2022fine,vu2021learned,hori2023learned,yu2015geospark,sabek2017spatial,you2015large,sedona,garcia2020efficient,jacox2007spatial}: global partitioning and local join. In the global partitioning phase, spatial datasets are divided into multiple partition blocks based on their spatial distribution, with denser regions receiving a greater number of partition blocks to balance the workload. During the local join phase, each worker is assigned one or more of these blocks. These workers then execute the join on their designated blocks. After each worker completes their tasks, the results are aggregated and refined to produce the final outcome. 

The mainstream systems for executing distributed spatial joins primarily utilize Hadoop~\cite{dean2008mapreduce} or Spark~\cite{zaharia2010spark} frameworks. These systems generally follow the two-phase approach, i.e., global partitioning and local join. Hadoop-based systems convert distributed spatial joins into a series of Map and Reduce tasks. Examples of Hadoop-based systems include \textit{Spatial-Hadoop}~\cite{eldawy2015spatialhadoop} and \textit{Hadoop-GIS}~\cite{aji2013hadoop}. However, a key limitation of Hadoop-based systems is their reliance on inefficient storage of intermediate results. To address this issue, Spark-based systems utilize Resilient Distributed Datasets (RDDs) and leverage in-memory processing to enhance efficiency. Examples of such include \textit{Apache Sedona}~\cite{sedona} (previously known as \textit{Geospark}~\cite{yu2015geospark}), \textit{Beast}~\cite{eldawy2021beast}, \textit{Simba}~\cite{xie2016simba}, \textit{LocationSpark}~\cite{tang2016locationspark,tang2020locationspark} and etc. In general, Spark-based systems are more effective due to their in-memory processing capabilities. 

\textbf{Machine Learning in Non-Spatial Joins:} Applying machine learning to non-spatial join operations lies in two primary categories. The first category involves the use of reinforcement learning to optimize the order of joins~\cite{krishnan2018learning,marcus2018deep,yu2020reinforcement,yan2023join,chen2022efficient}. In this approach, the system models the state as a specific configuration of join orders. Actions are defined as the potential joins that can be executed next, and rewards are inversely proportional to the query cost. The reinforcement learning model is trained to develop a policy that determines the most efficient join order, minimizing the overall query execution time. The second category involves the use of learned indexes to replace traditional components typically required in join operations~\cite{sabek2021case}. For example, Recursive Model Indexes (RMI)~\cite{kraska2018case} is used to replace conventional indexing mechanisms in Indexed Nested Loop Joins (INLJ). Similarly, CDF-based partitioning functions~\cite{kristo2020case} replace hash functions in hash-based joins (HJ), enhancing efficiency and performance. However, this idea cannot be applied to spatial join because spatial data requires specialized indexing and partitioning techniques.

\textbf{Machine Learning in Spatial Joins:}
Recent work  have focused on enhancing the efficiency of spatial join through learning-based methods. Vu et al.~\cite{vu2020using,vu2024learning} introduce a learning-based framework named \textit{Spatial Join Machine Learning (SJML)}, aimed at improving spatial join executions by utilizing statistics from the datasets involved. This framework offers a range of models, from selectivity estimation to selecting the appropriate join algorithms. However, some statistics, e.g., convoluted histogram~\cite{vu2020using,vu2024learning}, require traversing the datasets twice in real-time. This requirement degrades performance due to increased computational overhead.
In addition, there has been some work on the use of machine learning to identify effective partitioners. Hori et al.\cite{hori2023learned} propose a reinforcement learning model to achieve load-balanced partitioning for spatial joins. Their model conceptualizes the environment as two-dimensional grids, with actions defined as splitting a grid, and rewards measured by the runtime improvement from executing the join queries. However, this method is constrained by its requirement to perform numerous actual joins to evaluate rewards, which is time-consuming. Meanwhile, Vu et al.\cite{vu2020using} have developed a deep learning model that helps to select the best partitioning technique for spatial joins. By training on various data distributions, the model learns to correlate the characteristics of the data with the effectiveness of different partitioning techniques. 
However, it still encounters the overhead of partitioning data on-the-fly at runtime.

In contrast, our work is fundamentally different in that it uses machine learning not to select a partitioning algorithm, but to reuse an existing partitioner.
To the best of our knowledge, no previous work has utilized a machine learning approach to learn similarities between spatial join queries based on their inherent data characteristics and query embeddings. Our framework leverages these learned similarities to optimize new join executions by reusing partitioners from similar past queries, avoiding the need to construct a partitioner from scratch. This capability to generalize across different join configurations significantly improves performance, as demonstrated in our experiments.

\section{Preliminaries and Problem Definition}
\label{problem}

In this section, we formally define the terminologies used in this paper and present the problem we address.

\subsection{Preliminaries}
\textbf{Spatial data} refers to data associated with geographical locations, denoted as \( o \) (e.g., the coordinates that mark the location of a restaurant). A \textbf{spatial dataset} \( S \) consists of multiple spatial data records, such as a dataset that contains all the geographical locations of restaurants within a city, denoted as \( S = \{o_1, o_2, \ldots, o_n\} \).

\textbf{Spatial join} is a database operation that identifies pairs of objects from two spatial datasets that satisfy a specific spatial predicate \( \theta \), e.g., within a specific distance from each other. For example, a spatial join can be used to combine hotel and restaurant datasets to identify pairs that are within 500 meters of each other. Formally, given two spatial datasets \( R \) and \( S \), and a spatial distance \( \theta \), a spatial theta join between \( R \) and \( S \) based on \( \theta \) returns a set \( T \), where
 $$ T = \{\langle o_r, o_s \rangle | o_r \in R, o_s \in S, distance(o_r, o_s) \leq \theta \} $$
There are four fundamental spatial join algorithms~\cite{vu2021learned,vu2024learning}, \textit{Block Nested Loop Join (BNLJ)} ~\cite{belussi2020cost}, \textit{Distributed Join with Index (DJ)}~\cite{eldawy2015spatialhadoop,yu2015geospark}, \textit{Repartition Join (REPJ)}~\cite{eldawy2015spatialhadoop,yu2015geospark}, and \textit{Partition based Spatial Merge Join (PBSM)} ~\cite{patel1996partition}. \textit{BNLJ}  is the most straightforward approach, where spatial datasets are divided into blocks. For each block from the first dataset, all blocks from the second dataset are loaded and processed using a nested loop method. \textit{BNLJ} is efficient only in smaller datasets as it does not incorporate spatial properties for filtering~\cite{belussi2020cost}.
In contrast, \textit{DJ} shares a fundamental similarity to \textit{BNLJ} in its block-based processing approach but improves on it by utilizing spatial indexes. In \textit{DJ}, datasets are indexed, and objects closely located are stored within the same partition block, which facilitates the matching of overlapping blocks during partitioning. \textit{REPJ} addresses a critical limitation of \textit{DJ} that occurs when the indices from both datasets differ significantly in structure due to variations in data distributions and index types. This disparity can lead to inefficient matching. 
To overcome this, \textit{REPJ} uses the index of one dataset to repartition the other dataset, ensuring that the geometries overlapping from both datasets are located in the same partition block. \textit{PBSM} does not require any indices on the joining attributes of either dataset. It partitions both inputs into manageable chunks and employs a plane-sweeping technique to join these chunks efficiently.

\begin{figure*}
  \centering
  \includegraphics[width=\linewidth]{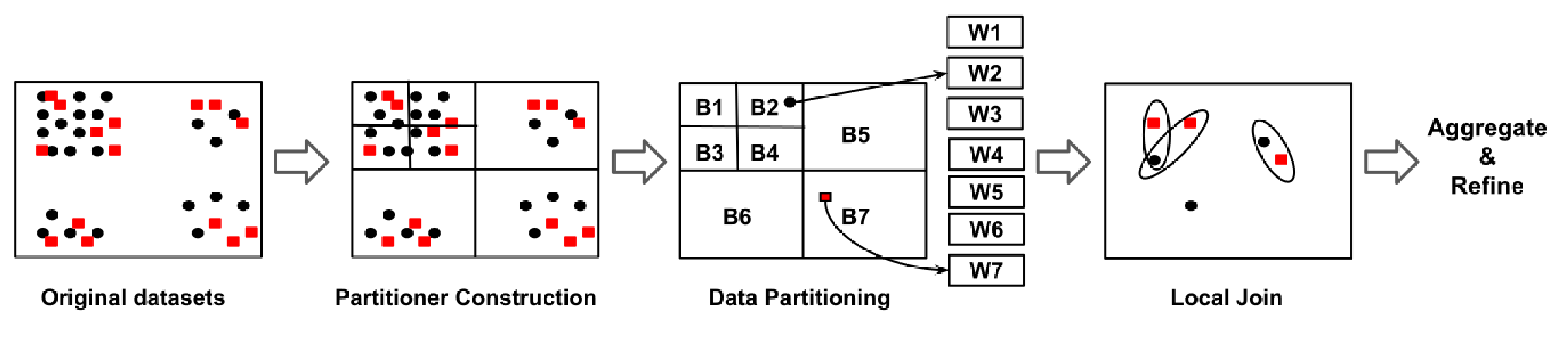}
  \vspace{-7mm}
  \caption{Overall Execution Flow of Distributed Spatial Join}
  \vspace{-3mm}
  \label{fig:joinoverall}
  
\end{figure*}

\textbf{Distributed spatial join} refers to the execution of a spatial join across multiple nodes within a distributed system 
As described in Section~\ref{relatedwork}, a distributed spatial join typically follows a two-step process: global partitioning and local join. Figure~\ref{fig:joinoverall} illustrates this process: two join datasets, represented by black circles and red rectangles, are initially divided by a partitioner into partition blocks such as B2 and B7, reflecting their data distribution. The data is then routed to different blocks according to their geographical location. The blocks are processed by the corresponding worker nodes, such as W2 and W7, to perform local joins. The results of each node are subsequently aggregated and refined.

\subsection{Problem Statement}
Our objective is to optimize distributed spatial joins by reducing distributed partitioning overhead.
We accomplish this by reusing existing partitioners and learning the similarities between different spatial join configurations. To this end, we introduce a problem named \textbf{Similarity-based Distributed Spatial Join (\textit{SDSJ})}.

The \textbf{input} of \textit{SDSJ} comprises a series of historical spatial join queries, denoted $\mathcal{J} = \{ J_1, J_2, \dots, J_n \}$. Each join query, represented as $J_i$, involves a pair of spatial datasets, $R_i$ and $S_i$, linked by a spatial predicate $\theta_i$. For each spatial join $J_i$, the system utilizes a specific partitioner $P_i$. This partitioner is typically associated with one of the datasets involved in the join—either $R_i$ or $S_i$—and co-partitions the other dataset. Each partitioner $P_i$ used is subsequently stored in a repository for potential reuse in future queries. 

The \textbf{output} is a machine learning model $M$ that learns the similarities between datasets using datasets embedding. For a new spatial join query, this model matches the datasets involved in the new join to existing ones in the repository and recommends an existing partitioner to execute the new join efficiently. Given a new spatial join query $J = (R, S)$, the model $M$ operates as follows:

\begin{itemize} [leftmargin=10pt, labelsep=.5em]
\item \textbf{Determine Maximum Similarity}: Identify the highest similarity score between $R$ and all datasets in $\mathcal{D}$ and between $S$ and all datasets in $\mathcal{D}$:
 $$ sim_{\text{max}} = \max (\max_{D_i \in \mathcal{D}} sim(R, D_i), \max_{D_i \in \mathcal{D}} sim(S, D_i)) $$

This maximum value represents the most similar existing dataset to $S$ and $R$, respectively.

\item \textbf{Partitioner Reuse Decision}: Evaluate whether $sim_{\text{max}}$ indicates sufficient similarity for partitioner reuse, if so, retrieve the partitioner $P_k$ associated with the dataset $D_k$ corresponding to $ sim_{\text{max}}$. Otherwise, partition the dataset on-the-fly (repartition).

\item \textbf{Join Execution}: Use the retrieved partitioner $P_k$ or a newly constructed partitioner to partition both $R$ and $S$.
Perform the spatial join operation using these partitions, leveraging the existing partitioning scheme to improve efficiency.

\end{itemize}

\section{Spatial Data Partitioning}
\label{partitioner}
Without loss of generality, we realize our techniques on top of Apache Sedona~\cite{sedona}, formerly known as GeoSpark~\cite{yu2019spatial}, as a major open-source distributed system for processing big spatial data extending Apache Spark.
Apache Sedona offers three partitioning algorithms for spatial join, namely uniform grid, quadtree~\cite{samet1984quadtree}, and KDB-tree partitioners~\cite{robinson1981kdb}.   The uniform grid partitioner divides the spatial universe into a fixed grid of uniform cells, independent of the data distribution. While it is simple, it typically leads to an imbalance in the number of data points per partition due to the typical skew in the spatial distribution. This imbalance significantly overloads certain worker nodes and reduces overall efficiency.

The KDB-tree partitioner divides the spatial domain by recursively splitting it based on the data distribution. However, its partitioning result can be sensitive to the sequence in which data points are inserted. This dependency on the insertion sequence introduces instability, meaning that different runs over the same dataset may produce different partitions. 

The quadtree partitioner recursively subdivides the spatial domain and produces consistent partitioning regardless of the insertion sequence. This consistency in partitioning makes it ideal for our purposes, as we need a partitioner whose outcome can be reliably reused across different datasets with similar characteristics. Given its advantages of consistency and balanced partitioning, we employ the quadtree partitioner in \textit{SOLAR}. In the original implementation of Apache Sedona, the quadtree is constructed by scanning the dataset to determine the Minimum Bounding Rectangle (MBR) of the data points and then sampling a subset of data to guide growing the tree to a certain depth. However, the original implementation in Apache Sedona constructs the quadtree only over the minimum bounding rectangle (MBR) of the current dataset. This limited spatial coverage makes it unsuitable for reuse if new datasets extend beyond that MBR. We introduce the following modifications:

\begin{itemize} [leftmargin=10pt, labelsep=.5em]
\item \textbf{Full Spatial Coverage:} Instead of starting with a Minimum Bounding Rectangle (MBR) that tightly fits the data points, we begin partitioning from the entire world's geographic space. This ensures complete coverage, making the resulting partitions reusable for various datasets, regardless of individual spatial extents.

\item \textbf{Adaptive Depth Mechanism:} In the original implementation, the maximum depth of the quadtree is determined by the number of Resilient Distributed Dataset (RDD) partitions used in Apache Sedona, based on Spark RDDs. To improve flexibility, we introduce an adaptive mechanism in which the depth is dynamically set as the maximum of two values: the number of RDD partitions and a user-defined maximum depth. This ensures that the quadtree is deep enough to capture the data distribution.
\end{itemize}

The objective is to make partitioning general enough to support diverse workloads while being specific to capture the underlying spatial distribution. By default, the partitioner in Apache Sedona~\cite{sedona} partitions the left-side dataset of the join. For example, in a join between \textit{restaurant} and \textit{park}, the \textit{restaurant} dataset is partitioned, and its partitioner is also used to partition the \textit{park} dataset.
After executing the join, we store the structure of the quadtree partitioner—including the spatial boundaries and hierarchy of partition blocks on disk. By storing this partitioner structure, the system can efficiently reuse it for subsequent joins involving similar datasets. Reusing an existing partitioner enables the system to route incoming data points directly to appropriate partitions, avoiding the overhead of recomputing the partitioning scheme from scratch.
\section{Dataset Similarity Evaluation}
\label{sim_metric}
In this section, we present how to capture the distribution of spatial datasets using histogram-based methods and evalaute the similarties between different datasets based on their respective histograms.

\subsection{Dataset Statistics Collection}
\label{sec:histograms}
\begin{figure}[t]
  \centering
  \includegraphics[width=\linewidth]{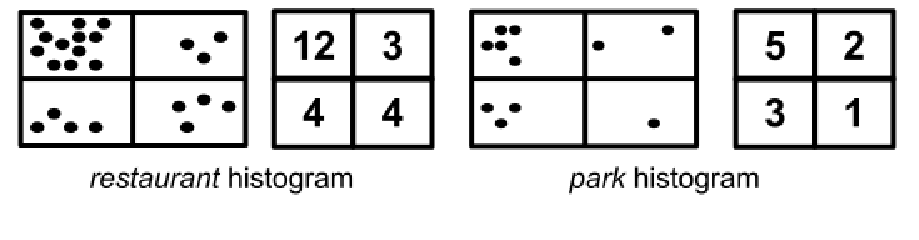}
  \vspace{-8mm}
  \caption{Data Histogram Construction}
  \vspace{-5mm}
  \label{fig:histogram}

\end{figure}

To effectively capture and analyze the data distribution of a dataset, we compute its histograms. A histogram is a statistical tool that groups data into bins; each bin represents a specific spatial range, and the count within each bin indicates the number of data points it contains. This method has been widely utilized in various studies~\cite{vu2020using,vu2021learned,vu2024learning,de2016distributed,das2004approximation} to analyze data distributions. Consequently, we compute data histograms for datasets to illustrate their data distributions. Each bin in the histogram counts the number of data points it contains, providing a detailed representation of the spatial distribution across the dataset. Figure~\ref{fig:histogram} shows the space division and construction of a 2X2 data histogram for the \textit{restaurant} and \textit{park} datasets. To utilize these histograms effectively in our Siamese Neural Network, we convert them into vector representations. This conversion involves flattening the histogram grid into a vector. Specifically, we concatenate the counts row by row from the histogram to form a large vector. In this example, the vector representation of the histogram corresponding to \textit{restaurant} is $[12, 3, 4, 4]$. In our experiment, we always use high-resolution histogram, i.e., 8192 X 8192, as suggested in ~\cite{vu2024learning}, due to its effectiveness in capturing fine-grained details of spatial distribution.

\subsection{JSD‐Based Similarity Measure}
\label{simeval}

We define the similarity between two datasets based on the distance between the vector representations of their data histograms that are computed in Section~\ref{sec:histograms}. To evaluate this distance, we use the Jensen-Shannon divergence ($JSD$)~\cite{lin1991divergence}, a measure of similarity between probability distributions.
In our case, $JSD$ is a highly effective measure to compare histograms. When it is computed with log to base 2, it provides values normalized in the range $[0,1]$. This bounded range makes model learning more effective because a normalized range allows for more efficient optimization during training. In addition, a similarity score within [0,1] is intuitively interpretable, where 0 indicates no similarity and 1 indicates identical distributions. This makes it easier to understand and analyze the model's outputs.
Unlike other distance metrics such as Euclidean distance, which only quantifies point-wise differences and is sensitive to the scale of data, Jensen-Shannon divergence captures the underlying distributional characteristics. It reflects both the shape and overlap of histograms, providing a more robust similarity metric that is less influenced by magnitude variations.  

A lower $JSD$ value indicates a smaller distance, i.e., a higher similarity, between the datasets. Before calculating the Jensen-Shannon divergence, the histogram vectors must be normalized to form probability distributions. This normalization is achieved by dividing each bin count by the total number of data points, transforming the vector into a probability distribution where each bin's value represents the probability of finding a data point in that bin.
The Jensen-Shannon divergence between the probability distribution vectors $\Tilde{H}_1$ and $\Tilde{H}_2$, corresponding to the original two histograms $H_1$ and $H_2$, is defined as follows:
\[
JSD(\Tilde{H}_1 \parallel \Tilde{H}_2) = \frac{1}{2} KLD(\Tilde{H}_1 \parallel M) + \frac{1}{2} KLD(\Tilde{H}_2 \parallel M)
\]
\[
M = \frac{1}{2}(\Tilde{H}_1 + \Tilde{H}_2)
\]
\[
KLD(\Tilde{H}_1 \parallel M) = \sum_{i} \Tilde{H}_1(i) \log\left(\frac{\Tilde{H}_1(i)}{M(i)}\right)
\]
\[
KLD(\Tilde{H}_2 \parallel M) = \sum_{i} \Tilde{H}_2(i) \log\left(\frac{\Tilde{H}_2(i)}{M(i)}\right)
\]
Where $M$ is the pointwise mean of the two probability distributions. Each Kullback-Leibler divergences ($KLD$)~\cite{kullback1951information} term calculates the divergence from one probability distribution to the mean, giving a measure of how much information is lost when $M$ is used to approximate $\Tilde{H}_1$ or $\Tilde{H}_2$. In the example from Figure~\ref{fig:histogram}, the original vectors corresponding to \textit{restaurant} dataset and \textit{library} dataset are $H_1 = [12,3,4,4]$ and $H_2 = [5,2,3,1]$, and their probability distribution vectors are $\Tilde{H}_1 = [0.522, 0.130, 0.174, 0.174]$ and $\Tilde{H}_2 = [0.455, 0.182, 0.273, 0.091]$. Then, the Jensen-Shannon divergence between $\Tilde{H}_1$ and $\Tilde{H}_2$ is calculated as follows. We first compute the pointwise mean vector $M = [0.488, 0.156, 0.223, 0.132]$ of $\Tilde{H}_1$ and $\Tilde{H}_2$. Next, we compute the Kullback-Leibler divergences, in our case $KLD(\Tilde{H}_1 \parallel M) = 0.0152$, and $KLD(\Tilde{H}_2 \parallel M) = 0.0156$. Finally, the Jensen-Shannon divergence is $JSD(\Tilde{H}_1 \parallel \Tilde{H}_2) = \frac{1}{2} \times  0.0152 + \frac{1}{2} \times 0.0156 = 0.0154$.

\section{\textit{SOLAR} Offline Phase}
\label{offline}
\textit{SOLAR} operates in two phases: offline and online. In the offline phase, it trains a model to capture similarities between datasets. Also, it evaluates some joins to train the model to decide whether a matched partitioner with a given similarity is similar enough for the new join execution. The online phase uses this model to process new join queries by reusing a suitable partitioner from the repository or creating a new one as needed. 

In this section, we detail the offline phase. This phase includes several key components:(1)~dataset embedding, (2)~the approximation of similarity using a Siamese Neural Network, and (3)~decision model training,  each outlined in Algorithm~\ref{offlinecode} and detailed below. 

\begin{algorithm}
\vspace{-1mm}
\caption{\textit{SOLAR} Offline Phase}
\label{offlinecode}
\begin{algorithmic}[1]
\State \textbf{Input:} set of  datasets $\{D_1, D_2, ... D_n\}$, sets of training joins $\{J_1, J_2, ... J_n\}$
\State \textbf{Output:} learned models from offline phase
\vspace{1mm}

\State \textbf{Step 1: Dataset Embedding}
\For{each dataset $D_i$}
    \State $emb(D_i)\leftarrow$ Extract polygon features from the metadata
\EndFor
\vspace{2mm}

\State \textbf{Step 2: Dataset Similarity Learning}
\State Initialize Siamese Neural Network with twin branches
\For{each pair of embeddings $emb(S), emb(R)$}
    \State $F(emb(S)), F(emb(R)) \leftarrow$ $emb(S), emb(R)$ in feature space
    \State Compute distance $ d \leftarrow \|F(emb(S)) -  F(emb(R)\|_2$
    \State Distance clamp $\hat{d} \leftarrow \frac{d}{1+d}$
    \State Compute MSE loss: $\|\hat{d} - d_{JSD} \|_2^2$
    \State Update network weights to minimize loss
\EndFor

\vspace{2mm}
\State \textbf{Step 3: Partitioner Reuse Decision Model}
\State Initialize empty sets $\mathcal{X}$, $\mathcal{Y}$ 
\For{each join $J_i$ in the training set}
    \State $sim_{max} \leftarrow$ Identify the max similarity using the network
    \State  $t_1  \leftarrow$  Reusing the best existing partitioner run the new join
    \State  $t_2  \leftarrow$  Building a new partitioner to run the join
    \State $label \leftarrow 1$ if $t1 < t2$ else 0

    \State Append $\bigl(sim_{max}, label\bigr)$ to $(\mathcal{X}, \mathcal{Y})$
\EndFor
\State Train a Random Forest using $\mathcal{X}$ (similarity scores) as features and $\mathcal{Y}$ (reuse a partitioner vs.\ build a new one) as labels

\end{algorithmic}
\end{algorithm}

\subsection{Dataset Embedding}
\label{embedding}
We have discussed utilizing Jensen-Shannon divergence to evaluate dataset similarities based on their data histograms. However, our ultimate objective in the online phase is to use a learned model to directly evaluate the similarities between different datasets using only the dataset embeddings, bypassing the expensive step of histogram construction. This will be facilitated by learning a Siamese Neural Network, as Section~\ref{simlearn} details. Consequently, it is important to develop a mechanism to effectively embed these datasets.

To achieve this, we design an encoding mechanism that leverages the metadata associated with each dataset, specifically the size of the dataset and its polygon covering. Our approach involves extracting meaningful information from the polygon covering of the spatial dataset, as shown in lines 3-6 in Algorithm~\ref{offlinecode}. The embedding mechanism encodes each spatial dataset into a concise yet expressive numerical vector, capturing the essential polygonal characteristics necessary for accurate similarity learning. Specifically, from the polygonal covering of each data set, we extract several meaningful geometric features, including the \textit{number of points}, \textit{polygonal area}, \textit{polygon centroid coordinates} ($centroid_x$, $centroid_y$),  \textit{bounding box} ($minx$, $miny$, $maxx$, $maxy$), and the \textit{compactness} measure, defined as $(4 \pi \times area)/(perimeter^2)$, to form a 9-dimensional vector.

\begin{figure}[t]
  \centering
  \includegraphics[width=.85\linewidth]{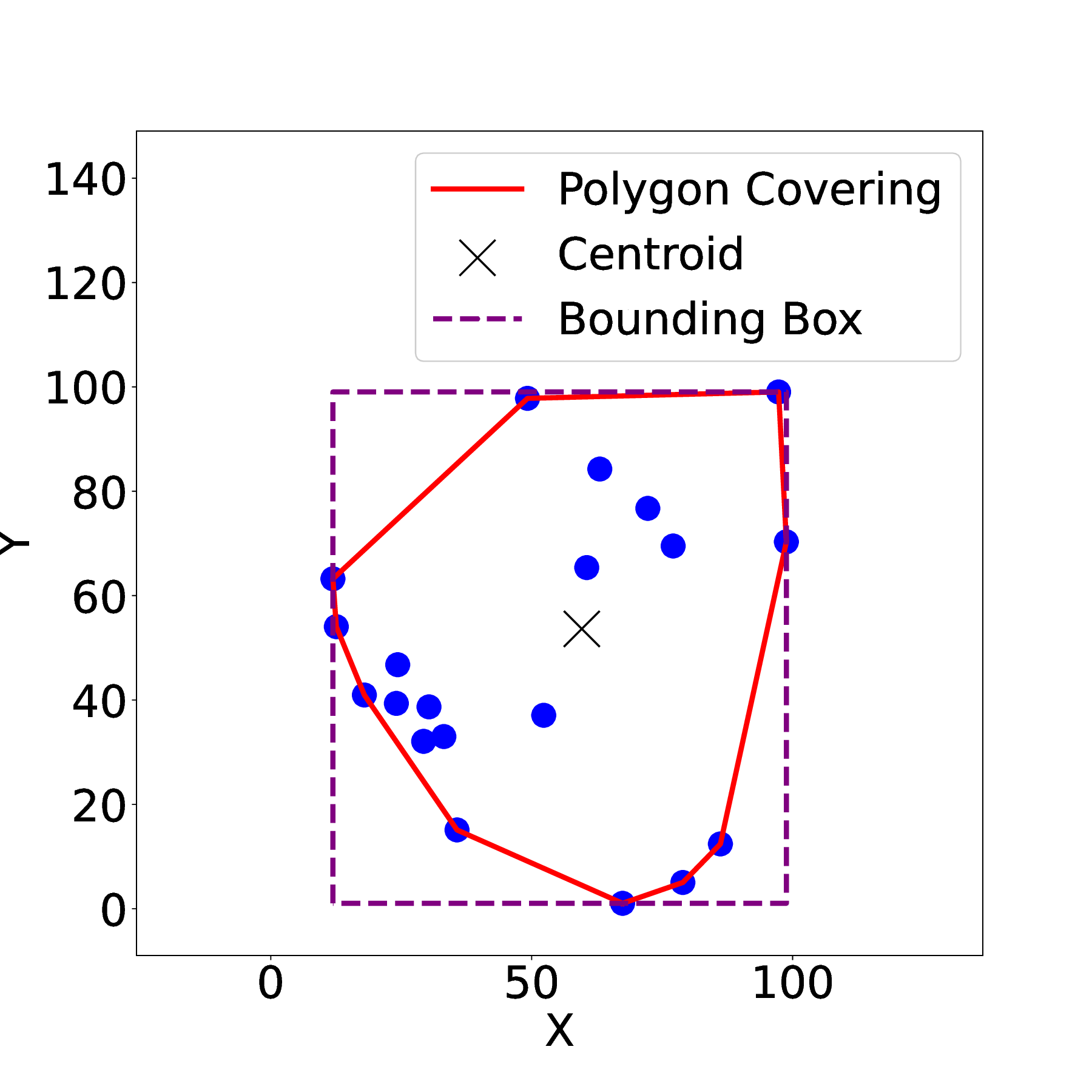}
  \vspace{-4mm}
  \caption{Dataset Embedding}
  \label{fig:encoding} 
  \vspace{-6mm}
\end{figure}

These geometric attributes concisely capture a dataset’s shape, spatial extent, and density. The bounding box and centroid convey location, the size and area track data density, and the compactness captures shape regularity. Moreover, by limiting ourselves to nine dimensions, we keep the embedding space small and efficient for our Siamese Neural Network training.
Figure~\ref{fig:encoding} gives an example of a spatial dataset along with its polygon covering. In this example, the number of points is $20$, the area of the polygon is $6196.79$, the centroid of the polygons is $(59.60, 53.62)$, the bounding box is $(11.90, 1.04, 98.81, 99.04)$, and the compactness is $0.87$.

To further improve the quality of the embeddings for effective learning, we apply normalization. Specifically, logarithmic scaling is employed for numerical stability and normalization of the polygon area and the number of points, while spatial coordinates under the CRS projection system, i.e.,  a standard way of specifying how map coordinates relate to real locations on Earth, are scaled down by a factor of $10^6$ to ensure consistent numerical magnitudes. Such normalization enhances the convergence stability of the Siamese Neural Network training process.

\subsection{Dataset Similarity Learning} \label{simlearn}

\begin{figure*}
  \centering
  \includegraphics[width=\linewidth]{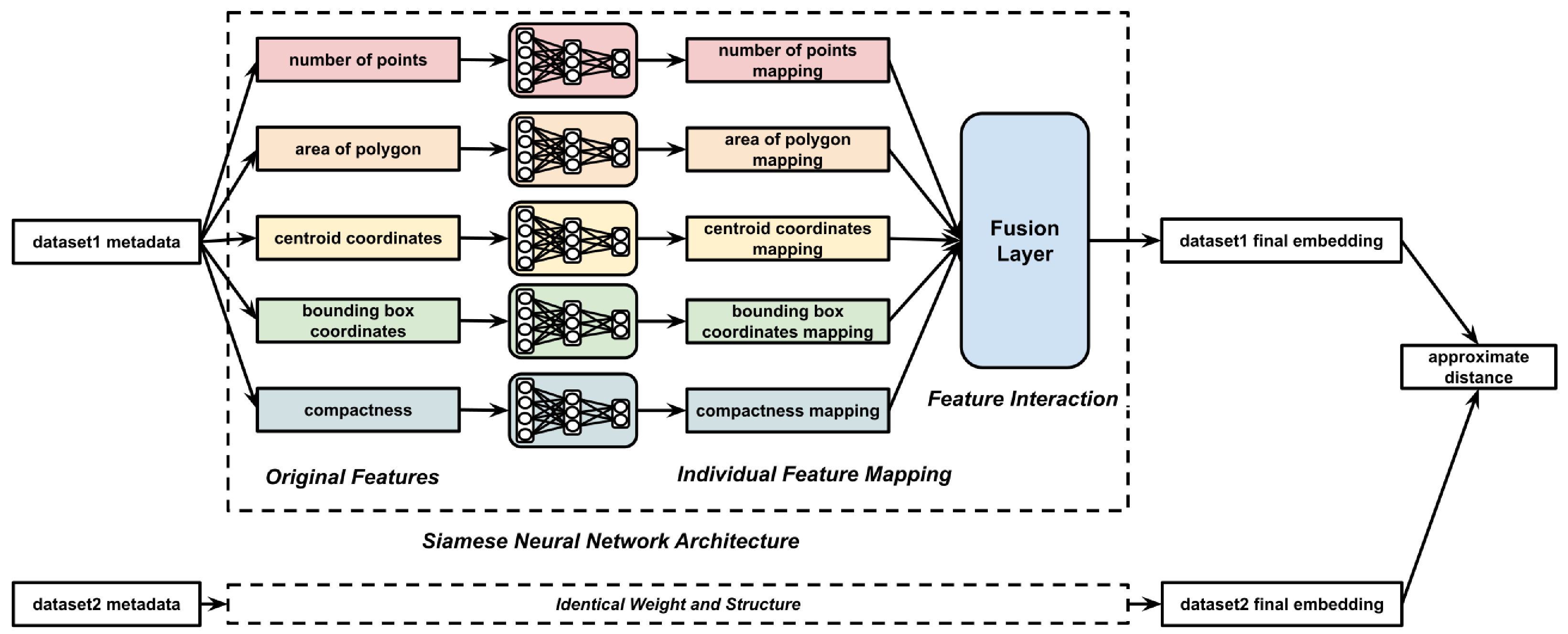}
  \vspace{-3mm}
  \caption{Siamese Neural Network}
  \vspace{-4mm}
  \label{fig:siamese}
\end{figure*}

In this section, we present our approach to learn the similarity between datasets using a Siamese Neural Network. This approach leverages the embeddings of datasets based on dataset metadata, i.e., polygon covering, to approximate the similarity between datasets.

\subsubsection{\textbf{Overview of Siamese Neural Networks}}
Siamese Neural Networks~\cite{bromley1993signature} are designed for tasks that involve comparing pairs of inputs to determine their similarity. In our framework, they are essential for identifying a similar dataset in the repository, allowing us to reuse its partitioner rather than recomputing one for each spatial join query. It works by computing the distance of pairs of input embeddings in the feature space, where a larger distance means less similarity and vice versa. These networks consist of two branches that share the same architecture and weights.
In our context, the inputs to the Siamese Network are the embeddings of two different datasets, denoted $emb(S)$ and $emb(R)$. Each branch of the network processes one of these embeddings. Denote the transformed encoding of $emb(S)$ and $emb(R)$ in the feature space as $F(emb(S))$ and $F(emb(R))$. Then we compute the Euclidean distance $d$ between their embedding in the feature space, i.e., 
$$ d = ||F(emb(S)) - F(emb(R))||_2$$
Furthermore, since the Jensen-Shannon divergence is bounded to the range $[0,1)$, we apply a clamping function to map $d$ into the $[0,1)$ interval:

$$\hat{d} = \frac{d}{1 + d}$$

Finally, $\hat{d}$ will be the predication of the distance between dataset $S$ and $R$ given by the Siamese Neural Network, which corresponds to the approximation of the ground truth distance between the two datasets based on the data histogram.

One notable property of the Siamese Neural Network is that if the same dataset reappears—meaning its metadata embedding is unchanged—the distance in the feature space is zero. This ensures the model will reliably retrieve the correct partitioner for any previously encountered dataset. As many enterprise workloads include repeated queries on largely overlapping datasets, this property can significantly improve performance in practical settings.

\subsubsection{\textbf{Loss Function}}
\label{lossandtraining}

To guide the network in learning a representation aligned with the Jensen--Shannon distance ($JSD$), 
we minimize a mean squared error loss between the predicted distance and the actual Jensen-Shannon Divergence score, specifically,

$$\mathcal{L} = \|\hat{d} - d_{JSD} \|_2^2$$

where $\hat{d}$ is the predicted distance given by the network, as illustrated above, and $d_{JSD}$ is the groundtruth distance computed according to the $JSD$ measure between data histograms.

\subsubsection{\textbf{Model Architecture}}
\label{modelarchitecture}
Our Siamese Neural Network is based on a modular design motivated by the intuition that different dataset features carry distinct semantic meanings and varying degrees of complexity.
Thus, processing these features individually allows the network to capture their unique characteristics more effectively. Specifically, we partition each dataset's 9-dimensional metadata into five groups: number of points, area, centroid, bounding box, and compactness. Each group is processed independently through its dedicated branch, implemented as specialized multi-layer perceptions.
For example, scalar features like the number of points and area are processed by simpler networks due to their limited complexity, whereas spatially informative features such as centroid and bounding box coordinates—which inherently contain richer spatial relationships—are handled by deeper subnetworks. Given their varying semantic meanings and complexity, each feature group is embedded independently through dedicated multi-layer perceptrons (MLPs), denoted as follows:

\begin{itemize}[leftmargin=12pt, labelsep=.5em]
    \item \textbf{Number of points} ($n_X$) is processed via:
    \[
    \begin{aligned}
    E_{A}(n_X) &= \text{ReLU}\Bigl(
        W^{(A)}_2 \,\text{ReLU}\bigl(W^{(A)}_1 n_X + b^{(A)}_1\bigr)
        + b^{(A)}_2\Bigr)
    \end{aligned}
    \]

    \item \textbf{Area of the Polygon} ($a_X$) is encoded similarly through:
    \[
    \begin{aligned}
    E_{B}(a_X) &= \text{ReLU}\Bigl(
        W^{(B)}_2 \,\text{ReLU}\bigl(W^{(B)}_1 a_X+ b^{(B)}_1\bigr) + b^{(B)}_2\Bigr)
    \end{aligned}
    \]

    \item \textbf{Centroid} $(c_{X_x}, c_{X_y})$ is processed through a deeper subnetwork:
    \[
    \begin{aligned}
    E_{C}(c_{X_x}, c_{X_y}) &= \text{ReLU}\Bigl(
        W^{(C)}_2 \,\text{ReLU}\bigl(W^{(C)}_1 [c_{X_x}, c_{X_y}]^T + b^{(C)}_1\bigr) + b^{(C)}_2\Bigr)
    \end{aligned}
    \]

    \item \textbf{Bounding Box} $(b_{X_x^{min}}, b_{X_y^{min}}, b_{X_x^{max}}, b_{X_y^{max}})$ is processed by:
    \[
    \begin{aligned}
    E_{D}(b_X) &= \text{ReLU}\Bigl(
        W^{(D)}_2 \,\text{ReLU}\bigl(W^{(D)}_1 [b_{X_x^{min}}, b_{X_y^{min}}, b_{X_x^{max}}, b_{X_y^{max}}]^T \\
        &\quad + b^{(D)}_1\bigr) + b^{(D)}_2\Bigr)
    \end{aligned}
    \]

    \item \textbf{Compactness} ($comp_X$) is embedded via:
    \[
    \begin{aligned}
    E_{E}(comp_X) &= \text{ReLU}\Bigl(
        W^{(E)}_2 \,\text{ReLU}\bigl(W^{(E)}_1\,comp_X + b^{(E)}_1\bigr)  + b^{(E)}_2\Bigr)
    \end{aligned}
    \]
\end{itemize}

Here, each $W^{(*)}_i$ and $b^{(*)}_i$ represent learnable parameters (weights and biases) of their respective fully connected layers. After these subnetworks independently process each component, the resulting embeddings are concatenated into an integrated representation $E_{comb}(emb_X)$.

Finally, to produce a unified and compact embedding capturing interactions among different features, 
we introduce a \textit{fusion layer}:
\[
\begin{aligned}
F(emb_X) 
&= \text{ReLU}\Bigl(
    W^{(fusion)}_2 \,\text{ReLU}\bigl(
       W^{(fusion)}_1\,E_{comb}(emb_X) 
    \bigr) \\
&\quad + b^{(fusion)}_1 + b^{(fusion)}_2\Bigr).
\end{aligned}
\]

This fusion step merges the complementary information from each branch, producing a compact embedding 
that comprehensively encodes the dataset’s characteristics. Figure~\ref{fig:siamese} illustrates the 
network architecture.



\subsubsection{\textbf{Summary}}

The primary goal of this model is to learn a function that computes the distance between datasets using embeddings from dataset metadata to approximate the actual distances computed using histograms. A larger distance represents less similarity and vice versa. Notice that a simple partitioner lookup will work only for queries that have been seen before. Using the model works for new queries that are somewhat similar yet not identical to the previous queries. The pseudocode for training the Siamese Neural Network is given in lines 7-15 in Algorithm~\ref{offlinecode}.

\subsection{Partitioner Reuse Decision}
\label{decision-maker}
Section~\ref{simlearn} discussed using the Siamese Neural Network to estimate the distance between datasets. However, selecting the appropriate similarity threshold to determine whether to reuse an existing partitioner remains challenging, as theoretically modeling the complexity of spatial joins is a difficult optimization problem~\cite{vu2024learning}. To address this, in the offline phase, we execute a subset of join queries with and without reusing the matched partitioner, recording the respective runtimes. This empirical data allows us to train a lightweight classification model to decide, at runtime, whether partitioner reuse is likely to be beneficial.  We collect empirical data from a series of join experiments. Specifically, we record:

\begin{itemize}[leftmargin=12pt, labelsep=.5em]
  \item \emph{similarity score} for each dataset pair, as produced by the network.
  \item The runtime $t_1$ if we reuse an existing (matched) partitioner.
  \item The runtime $t_2$ if we construct a new partitioner on-the-fly.
\end{itemize}

In principle, reusing a partitioner is beneficial if $t_1 < t_2$. However, if the best match in our repository is only weakly similar to the current dataset, the reused partitioner might be suboptimal or even cause the join to fail due to heavily skewed partitions. We treat this as a classification task: given the similarity score, we predict whether reusing the existing partitioner with the given best similarity score will be faster (i.e., $t_1 < t_2$) or not.

We choose a random forest classifier~\cite{breiman2001random} because it is effective in capturing complex decision surfaces via an ensemble of decision trees, while inherently mitigating overfitting through bagging. In our use case, this means we can accommodate cases where a small increase in similarity might drastically alter partition‐reuse performance for certain dataset pairs. We label each training instance as 1 if \(t_1 < t_2\), and 0 otherwise. We then train a random forest on these labeled samples, using only the \emph{similarity score}. The pseudocode of training the decision model is given in line 16-25 in Algorithm~\ref{offlinecode}. We call this the decision maker, which is used to determine whether the matched partitioner should be used to partition the new incoming join with the given similarity score.

\subsection{Model Maintenance}
The system’s data repository and workload evolve over time. Therefore, it is essential to maintain the learning model so that they continue to capture the distribution of new datasets. To address this, we adopt a strategy inspired by ~\cite{zhu2024pilotscope}, which advocates periodic and feedback-based retraining:

\begin{itemize}[leftmargin=12pt, labelsep=.5em]
    \item \textbf{Periodic Retraining:} We schedule regular retraining intervals (e.g., weekly or monthly, depending on the workload update rate), during which we collect all newly arrived datasets and any newly logged join feedback. At the end of each interval, we retrain the Siamese Neural Network with the expanded set of dataset embeddings, ensuring the embedding space reflects the broader data distribution. Once the updated Siamese model is validated, we also retrain the Random Forest decision model on the new similarity scores and observed runtime labels. 
    \item \textbf{Feedback-based Retraining:} Beyond fixed timers, the system administrator can monitor recent query executions—especially cases where reusing a partitioner yields suboptimal runtime or fails to maintain balanced partitions. If repeated anomalies are detected, we trigger an earlier retraining cycle. This enables the models to adapt promptly to abrupt changes in data distributions or the introduction of fundamentally new datasets.
\end{itemize}
\section{\textit{SOLAR} Online Phase} \label{online}

In the online phase, \textit{SOLAR} employs the trained Siamese Neural Network and the decision maker built in the offline phase to process incoming spatial join queries efficiently.
When a new join involving datasets $R$ and $S$ arrives, the system aims to leverage existing partitioners by identifying the most similar datasets from a repository $\mathcal{D}$. The procedure is outlined in Algorithm~\ref{onlinecode} as follows:

\begin{enumerate}[leftmargin=12pt, labelsep=.5em]

\item \textbf{Embedding Generation}: Compute the embeddings for the new datasets, $emb(R)$ and $emb(S)$, using the same embedding method applied during the offline phase. This initial step is captured in Algorithm~\ref{onlinecode} from lines 3 to 5.

\item \textbf{Similarity Computation}: For each dataset $D_i$ in the repository $\mathcal{D}$, compute the similarity between $R$ and $D_i$, denoted as $\text{sim}(R, D_i)$, and the similarity between $S$ and $D_i$, denoted as $\text{sim}(S, D_i)$, using the vectorized computation through the Siamese Neural Network, as shown in Algorithm~\ref{onlinecode} from lines 6 to 9. For each $R$ and $S$, find the maximum similarity values:
 $$sim_{max} = (\max_{D_i \in \mathcal{D}} \textit{sim}(R, D_i), \max_{D_i \in \mathcal{D}} \textit{sim}(S, D_i))$$
\item \textbf{Partitioner Reuse Decision}: Use the decision maker, as detailed in Section~\ref{decision-maker}, to determine whether to reuse the matched partitioner with the given similarity score. This decision-making is reflected in lines 10 to 12 of Algorithm~\ref{onlinecode}. If a partition reuse decision is not made, an on-the-fly partitioner is constructed.

\item \textbf{Join Execution}: Apply the partitioner to partition both $R$ and $S$, and execute the spatial join using these partitions. The metadata of the datasets is kept for subsequent training.

\end{enumerate}

\vspace{-3mm}
\begin{algorithm}
\caption{\textit{SOLAR} Online Phase}
\label{onlinecode}
\begin{algorithmic}[1]
\State \textbf{Input:} Join $J = (R, S)$, Repository $\mathcal{D}$, trained Siamese Neural Network, Threshold $\theta$
\State \textbf{Output:} Reused or newly created partitioner for $R$ and $S$
\vspace{1mm}

\State \textbf{Step 1: Embedding Generation}
\State $emb(R) \leftarrow$ Generate embedding for dataset $R$
\State $emb(S) \leftarrow$ Generate embedding for dataset $S$
\vspace{1mm}

\State \textbf{Step 2: Similarity Computation}
\State $sim_{\text{max}}(R) \leftarrow \max_{D_i \in \mathcal{D}} sim(R, D_i)$
\State $sim_{\text{max}}(S) \leftarrow \max_{D_i \in \mathcal{D}} sim(S, D_i)$
\State $sim_{\text{max}} \leftarrow \max (sim_{\text{max}}(R), sim_{\text{max}}(S))$
\vspace{1mm}

\State \textbf{Step 3: Partitioner Reuse Decision}
\State y = \textit{Decision Maker ($sim_{max}$)} 
\State $P_k \leftarrow$ matched partitioner if $y= 1$, else computed on-the-fly
\vspace{1mm}

\State \textbf{Step 4: Join Execution}
\State Apply $P_k$ to partition both $R$ and $S$
\State Execute the spatial join using these partitions

\end{algorithmic}
\end{algorithm}

\vspace{-2mm}
\section{Experimental Evaluation}
\label{experiment}

In this section, we present experimental evaluations to demonstrate the efficiency of \textit{SOLAR} in executing distributed spatial joins queries. Section~\ref{setup} introduces the experimental setup. Section~\ref{performanceeval} evaluates the performance of \textit{SOLAR} against the baselines. Our experiments aim to answer the following questions. 

\begin{itemize}[leftmargin=12pt, labelsep=.5em]
\item How often does \textit{SOLAR} reuse an existing partitioning scheme for new join queries?
\item Does the learning-based model introduce runtime overhead at online phase?
\item By how much can \textit{SOLAR} accelerate distributed spatial joins compared to the baselines? 
\item How does \textit{SOLAR} perform under different join predicates?

\end{itemize}

\subsection{Experimental Setup} \label{setup}

In this section, we outline the experimental settings, including the evaluation datasets, queries to evaluate, selection of the baseline algorithms, performance measures, parameter settings, and configurations for our computing clusters.

\textbf{Evaluation Datasets:} We collected a total of 37 datasets from multiple sources and across diverse geographic scales. i.e., from city-level to world-scale. These datasets include two traffic collision datasets with 1.7 million and 2.1 million data points, respectively, corresponding to vehicle collisions in New York City and Seattle \cite{NYCdata,Seattledata}, as well as a crime dataset containing 7.7 million data points from Chicago \cite{Chicagodata}. Additionally, we obtained a dataset from the Twitter API \cite{TwitterData2025} comprising 40 million data points and 33 datasets from OpenStreetMap \cite{osm}. Specifically, datasets from OpenStreetMap cover various points of interest such as shops, fire stations, and libraries. These datasets span three regions: China, the United States, and the entire world. To effectively test our system's performance on large datasets, the original datasets from OpenStreetMap are enlarged to sizes ranging between 5 million and 50 million data points. This enlargement is achieved by modeling the spatial distribution of the original data using a two-dimensional histogram and generating additional data points by sampling from this distribution. This approach ensures that the augmented datasets preserve the spatial characteristics of the original data. Our selection of datasets allows us to evaluate \textit{SOLAR} across different geographic scales and spatial distributions.  For our experiments, we split each dataset into disjoint \textit{training datasets} and \textit{test datasets} (by default, 80\% train and 20\% test). \textit{Training datasets} are used to train our spatial partitioning model. \textit{Test datasets} are used to build spatial joins that are run against \textit{SOLAR}. The objective is to reuse a similar previously seen partitioner whenever possible.

The datasets from the \textit{training datasets} are partitioned, and their corresponding partitioners are stored in our repository to train our model. We also form pairs of \textit{training datasets} to compute ground-truth similarity via histogram, as mentioned in Section~\ref{sim_metric}. \textit{test datasets} are unseen during any preprocessing or offline phases, thus testing the generalizability of the model. However, it is expected that a portion of the \textit{test datasets} share similar spatial properties and distribution to some of the \textit{training datasets}. This portion increases as the size of the \textit{training datasets} increases.

\label{exp-queries}
\textbf{Queries:} $SOLAR$ executes spatial distance joins between pairs of datasets.  We choose the query workload by randomly pairing the \textit{training datasets} to form \textit{training joins} such that every \textit{training dataset} appears in at least one join, and the total number of \textit{training joins} equals the number of \textit{training datasets}. Meanwhile, the \textit{test datasets} are also randomly paired to generate \textit{test joins}, which consist of entirely new dataset pairs that never appeared during training. By design, the \textit{training joins} and \textit{test joins} sets are disjoint, i.e., no \textit{training dataset} appears in a \textit{test join}. The total number of \textit{training joins} and the number of \textit{test joins} is 40 joins.

Notice that in our evaluation of \textit{SOLAR}, we follow these steps: (1) we train \textit{SOLAR} using joins from the training joins, and (2) we evaluate the performance by running queries against \textit{SOLAR}. These new queries can be either repeated queries, i.e., queries originating from the training joins, or new queries, i.e., queries originating from the test joins. When we evaluate queries from the training joins, we assess \textit{SOLAR}'s ability to detect repeated spatial joins and use a pre-existing spatial partitioner. When we evaluate queries from the test joins, we assess \textit{SOLAR}'s ability to evaluate the spatial similarity between new, unseen datasets that were not included in the training of \textit{SOLAR}. \textit{SOLAR} then needs to decide whether to reuse a similar pre-existing partitioner or partition the data on the fly for the upcoming join.

\textbf{Baseline Algorithms:}
We use two variants of \textit{Apache Sedona}~\cite{sedona} (previously known as Geospark~\cite{yu2015geospark}), the state-of-the-art open-source distributed system to process distributed spatial joins, as baselines against our \textit{SOLAR}. The two variants are denoted as \textit{Sedona-K}, which employs KDB tree as the partitioner, and \textit{Sedona-Q}, which uses Quadtree as the partitioner.  Notice that there are learning-based approaches from related work~\cite{hori2023learned,vu2020using}, but these approaches are not directly comparable to \textit{SOLAR}. First, \cite{hori2023learned} only supports self-joins and requires running an entire join repeatedly to compute a reward function, an impractical overhead for large datasets (e.g., tens of millions of points). Second, \cite{vu2020using} focuses only on deciding \emph{which} of Sedona’s standard partitioners to use. Our evaluation already compares \textit{SOLAR} against \textit{all} of Sedona’s built-in partitioners (including KDB-tree and Quadtree), so the functionality of \cite{vu2020using} is inherently covered as we compare \textit{SOLAR} against all possible partitioners of Sedona.

\textbf{Performance Measures:}
We execute spatial join queries (either from \textit{training joins} or \textit{test joins}) and analyze their runtime using several metrics. If \textit{SOLAR}’s decision maker opts to reuse a partitioner, the observed runtime is reported under \textit{SOLAR}. Otherwise, if reuse is found less efficient, \textit{SOLAR} constructs a new partitioner, reverting to the baseline.  We detail the frequency of this reuse in Section~\ref{reuse}. We report the best runtime, 25th percentile, median (50th percentile), 75th percentile, and worst runtime across all join executions. Since each join query varies in computational complexity due to differences in input sizes and spatial distributions, we further assess the performance of \textit{SOLAR} compared to the baseline \textit{Sedona} using speed-up ratio, defined as $\frac{runtime(Sedona)}{runtime(SOLAR)}$, where $runtime(Sedona)$ is the faster of  \textit{Sedona-K} and  \textit{Sedona-Q} and $runtime(SOLAR)$ is the runtime of \textit{SOLAR} in executing the query.

\textbf{Parameter Setting:} 
For the Siamese Neural Network, the features of each dataset are grouped into five subsets: (\textbf{A}) number of points, (\textbf{B}) area, (\textbf{C}) centroids, (\textbf{D}) bounding box, and (\textbf{E}) compactness. Subsets A, B, E (single-scalar inputs) each pass through a two-layer MLP with 8 and 4 hidden units per layer (ReLU), while C (2D) uses a two-layer MLP with 16 and 8 units, and D (4D) uses a deeper two-layer MLP with 32 and 16 units. The outputs of these subnetworks are concatenated into a 36-dimensional vector and fed into a two-layer fusion MLP (16 and then 8 hidden units, ReLU) to produce an 8-D embedding. We train with mean squared error (MSE) loss against the ground-truth Jensen-Shannon divergence. The training uses the Adam optimizer with a learning rate selected from ${0.0001, 0.0003, 0.001, 0.003, 0.01}$ and weight decay of $0$ and $0.0001$, chosen via 5-fold cross-validation. The final model is trained with the best hyperparameters on the full training set using a batch size of 24, for up to 50 epochs with early stopping (patience 10).  The random forest model used in the decision maker is constructed with 100 trees, each of maximum depth 5.

\textbf{Computing Cluster Configuration:} 
Our experiments are conducted on a Microsoft Azure cluster consisting of two head nodes (E8 V3, 8 cores, 64 GB RAM) and eight worker nodes (E2 V3, 2 cores, 16 GB RAM). All data is stored in HDFS, and we use Apache Sedona~\cite{sedona} in YARN mode to execute distributed spatial join queries.

\subsection{Performance Evaluation}
\label{performanceeval}
This section evaluates \textit{SOLAR} against the baseline algorithms, i.e., \textit{Sedona-Q} and \textit{Sedona-K}, under various configurations. 

\begin{figure}[t]
  \centering
  \includegraphics[width=.7\linewidth]{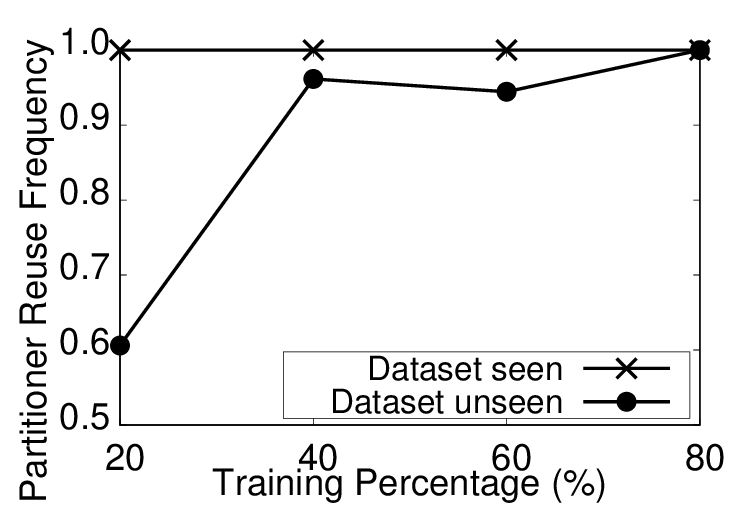}
  \caption{Matching Frequency Vs Training Data Percentage}
  \vspace{-3mm}
  \label{matching-frequency}
  \vspace{-1mm}
\end{figure}

\subsubsection{\textbf{Partitioner Reuse Frequency Analysis}}
\label{reuse}
Figure~\ref{matching-frequency} illustrates the frequency with which an existing partitioner is selected for incoming joins. We examine cases where upcoming joins are repeated (i.e., were seen during \textit{SOLAR}'s training) or not repeated (i.e., were not seen during \textit{SOLAR}'s training). In this experiment, we increase the percentage of datasets used in \textit{SOLAR}'s training from 20\% to 80\% of our overall spatial datasets.

The figure highlights that when \textit{SOLAR} is presented with incoming joins composed of datasets encountered during training, its model accurately identifies these as repeated operations and efficiently reemploys a pre-computed partitioner. This ability allows \textit{SOLAR} to effectively detect and bypass redundant partitioning for recurring join tasks. The reason is that identical datasets yield identical embeddings, resulting in a feature-space distance of zero, indicative of maximum similarity, as explained in Section~\ref{offline}.

However, if we build joins from \textit{test datasets} that have not been directly seen before during the training of \textit{SOLAR}, as we increase the number of \textit{training datasets} used int the training of \textit{SOLAR}, the number of corresponding \textit{training joins} grows proportionally. As a result, the system encounters and stores a larger variety of dataset pairs and their associated partitioners during the offline phase. Consequently, the probability of successfully matching a dataset from a new join to an existing partitioner increases significantly. This explains the upward trend observed in Figure~\ref{matching-frequency}, where the likelihood of partitioner reuse for new incoming joins improves with a larger set of \textit{training datasets} and \textit{training joins}. This further highlights the applicability and scalability of our architecture in industrial environments, where datasets are continuously integrated into production workflows. As the model accumulates more datasets, the repository of partitioners expands, thereby enhancing the system’s capability to effectively reuse existing partitioners.

\subsubsection{\textbf{Partitioning Phase Speed-up}}
Table~\ref{tab:partitiontime_improvement} shows the speed-ups achieved by the partitioning phase of \textit{SOLAR}. It shows that \textit{SOLAR} achieves up to 2.71X faster compared to the baselines. In contrast to the baseline approaches, which require two scans of the input data—first to collect data samples and calculate the minimum bounding rectangle (MBR) of the input for constructing the partitioner, followed by a second scan to route the data to partitions—\textit{SOLAR} leverages an existing partitioner to immediately direct data to the appropriate partitions.

\begin{table}[ht]
\centering
\vspace{-1mm}
\caption{Speed-up Ratio for Partitioning Phase}

\label{tab:partitiontime_improvement}
\begin{tabular}{@{}ccccccc@{}}
\toprule
\textbf{Case} &  \textbf{Worst} & \textbf{25th } & \textbf{50th} & \textbf{75th} & \textbf{Best} \\ \midrule
\multirow{1}{*}{train} 
 & 1.83  & 2.07 & 2.16 & 2.30 & 2.70 \\ \midrule
\multirow{1}{*}{test} 
 & 1.96 & 2.01 & 2.26 & 2.40 & 2.71 \\ \bottomrule
\end{tabular}
\vspace{-5mm}
\end{table}

\begin{figure*}
  \begin{subfigure}[b]{.41\columnwidth}
    \includegraphics[width=\textwidth]{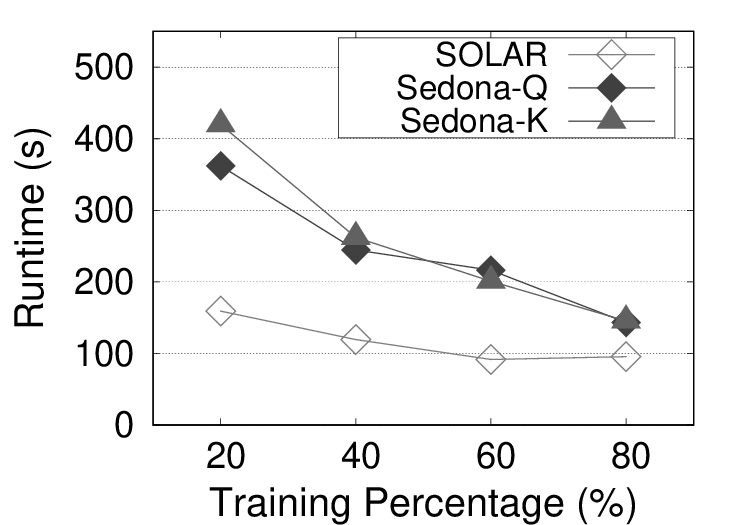}
    \caption{Best Case}
  \end{subfigure}
  \hfill
  \begin{subfigure}[b]{.41\columnwidth}
    \includegraphics[width=\textwidth]{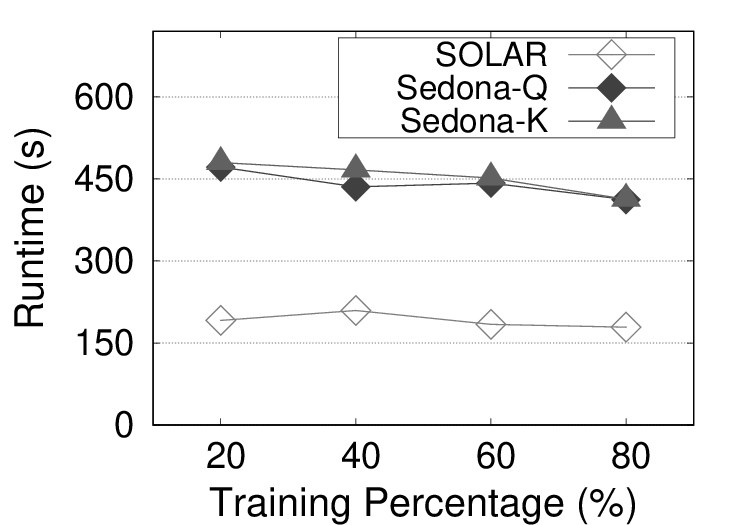}
    \caption{25 Percentile}
  \end{subfigure}
  \hfill
  \begin{subfigure}[b]{.41\columnwidth}
    \includegraphics[width=\textwidth]{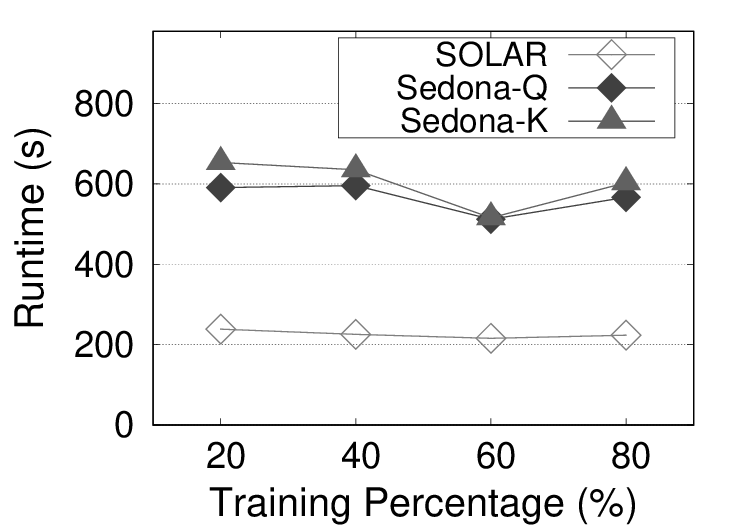}
    \caption{50 Percentile}
  \end{subfigure}
  \hfill
  \begin{subfigure}[b]{.41\columnwidth}
    \includegraphics[width=\textwidth]{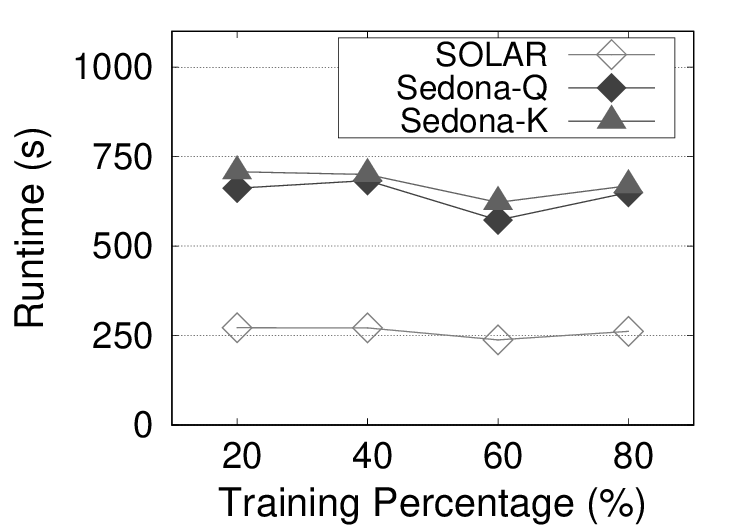}
    \caption{75 Percentile}
  \end{subfigure}
  \hfill
  \begin{subfigure}[b]{.41\columnwidth}
    \includegraphics[width=\textwidth]{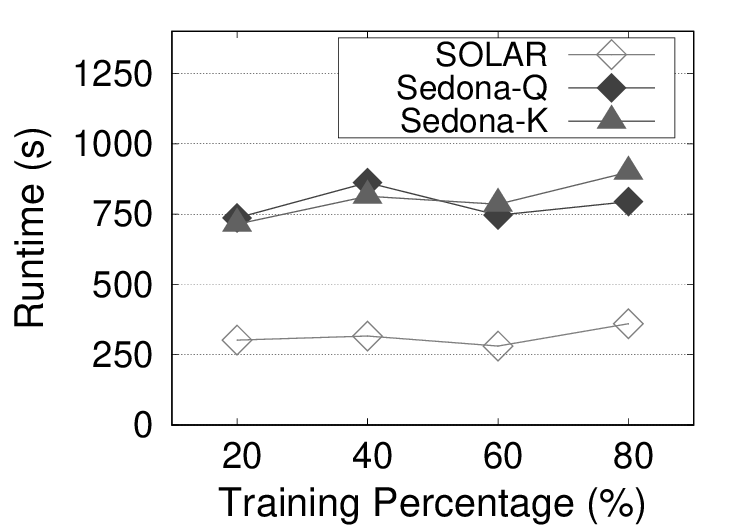}
    \caption{Worst Case}
  \end{subfigure}
  \vspace{-3mm}
  \caption{Runtime under Different Percentage of Training Data for \textit{training joins}}
  \vspace{-2mm}
  \label{vary-training-percentage-train}

\end{figure*}

\begin{figure*}
  \begin{subfigure}[b]{.41\columnwidth}
    \includegraphics[width=\textwidth]{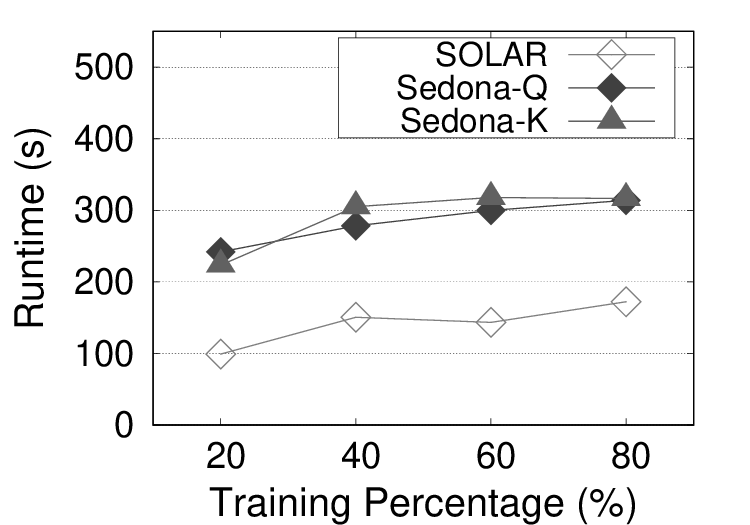}
    \caption{Best Case}
  \end{subfigure}
  \hfill
  \begin{subfigure}[b]{.41\columnwidth}
    \includegraphics[width=\textwidth]{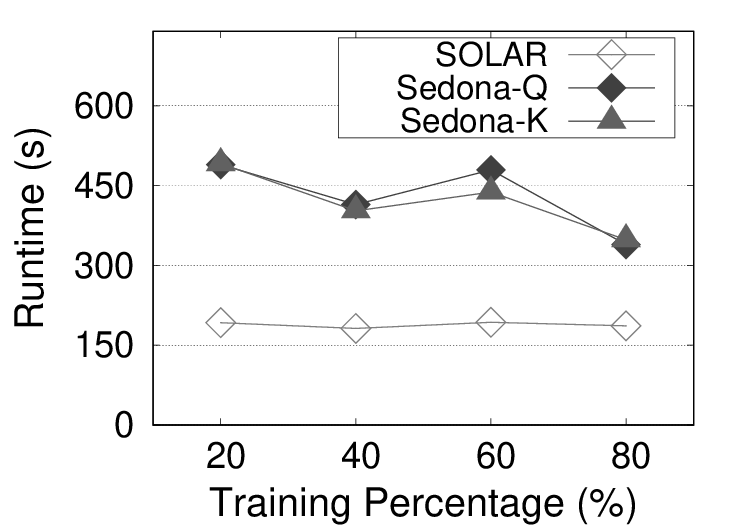}
    \caption{25 Percentile}
  \end{subfigure}
  \hfill
  \begin{subfigure}[b]{.41\columnwidth}
    \includegraphics[width=\textwidth]{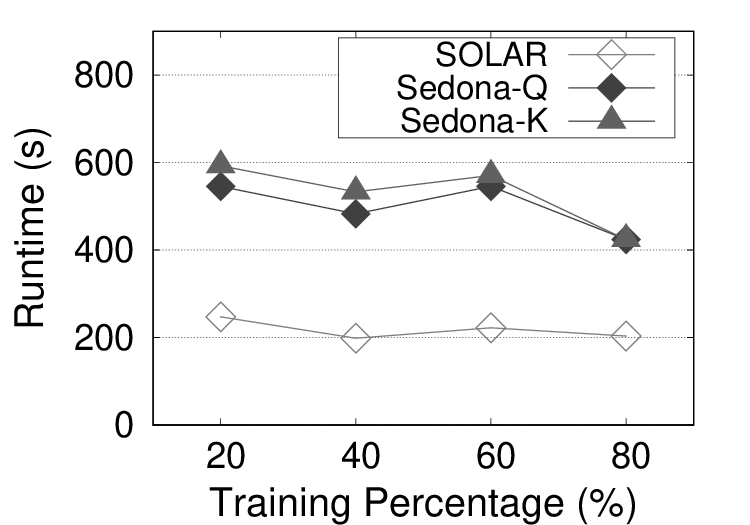}
    \caption{50 Percentile}
  \end{subfigure}
  \hfill
  \begin{subfigure}[b]{.41\columnwidth}
    \includegraphics[width=\textwidth]{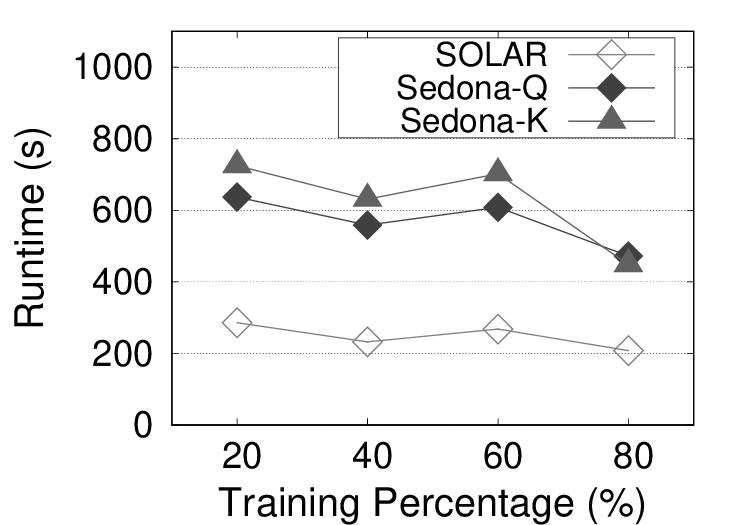}
    \caption{75 Percentile}
  \end{subfigure}
  \hfill
  \begin{subfigure}[b]{.41\columnwidth}
    \includegraphics[width=\textwidth]{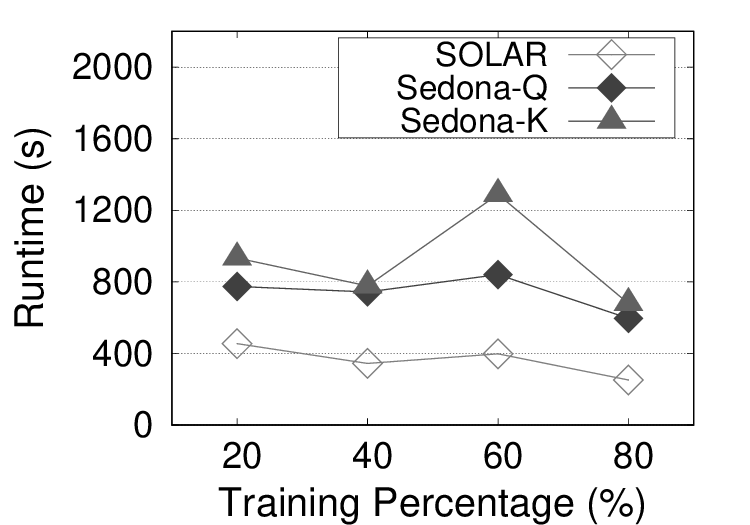}
    \caption{Worst Case}
  \end{subfigure}
  \vspace{-3mm}
  \caption{Runtime under Different Percentage of Training Data for \textit{test joins}}
  \vspace{-2mm}
  \label{vary-training-percentage-test}

\end{figure*}

\begin{figure*}
  \begin{subfigure}[b]{.41\columnwidth}
    \includegraphics[width=\textwidth]{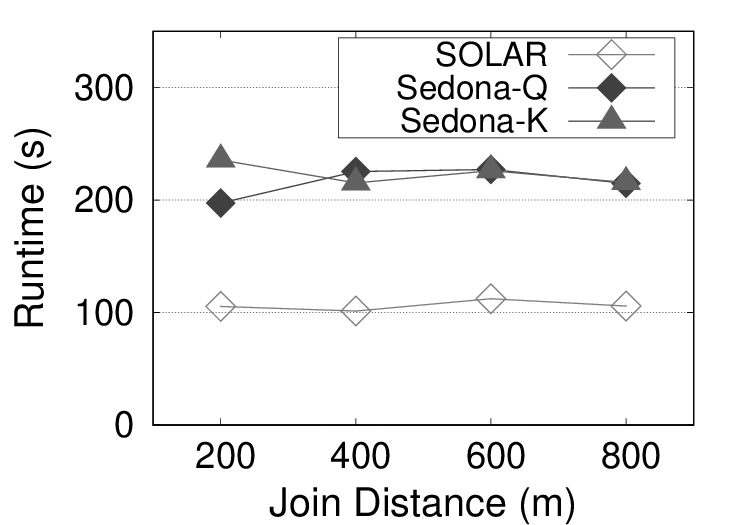}
    \caption{Best Case}
  \end{subfigure}
  \hfill
  \begin{subfigure}[b]{.41\columnwidth}
    \includegraphics[width=\textwidth]{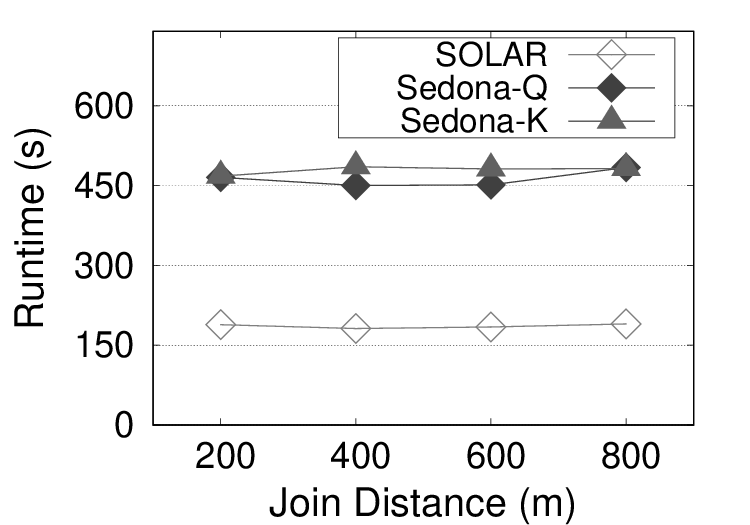}
    \caption{25 Percentile}
  \end{subfigure}
  \hfill
  \begin{subfigure}[b]{.41\columnwidth}
    \includegraphics[width=\textwidth]{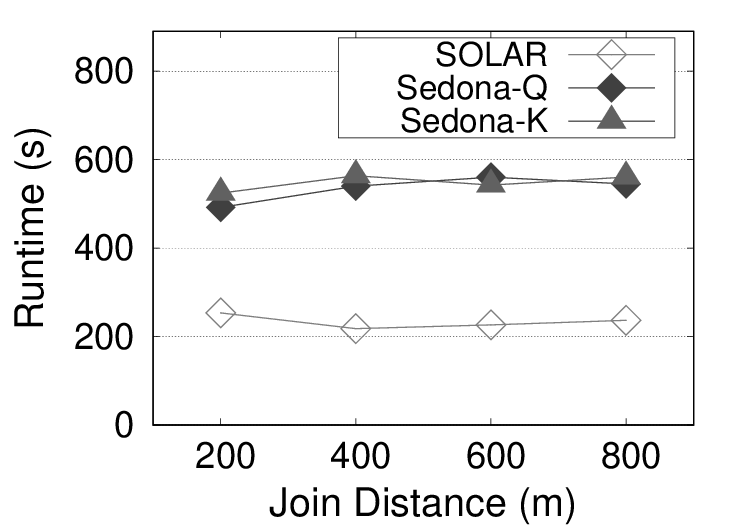}
    \caption{50 Percentile}
  \end{subfigure}
  \hfill
  \begin{subfigure}[b]{.41\columnwidth}
    \includegraphics[width=\textwidth]{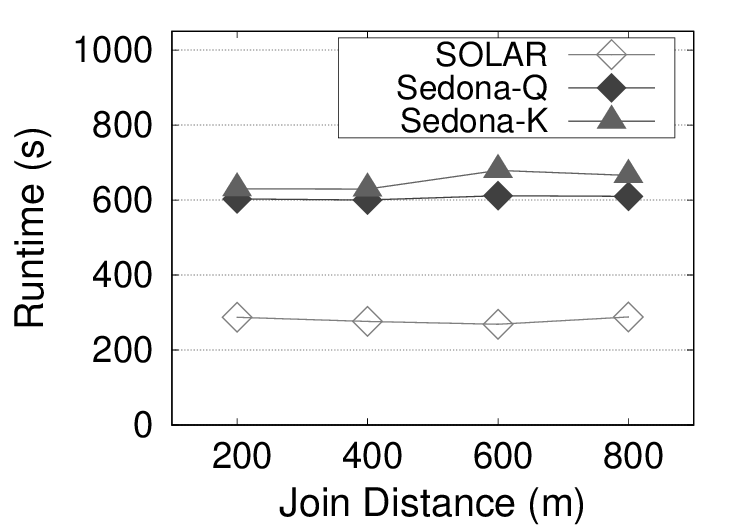}
    \caption{75 Percentile}
  \end{subfigure}
  \hfill
  \begin{subfigure}[b]{.41\columnwidth}
    \includegraphics[width=\textwidth]{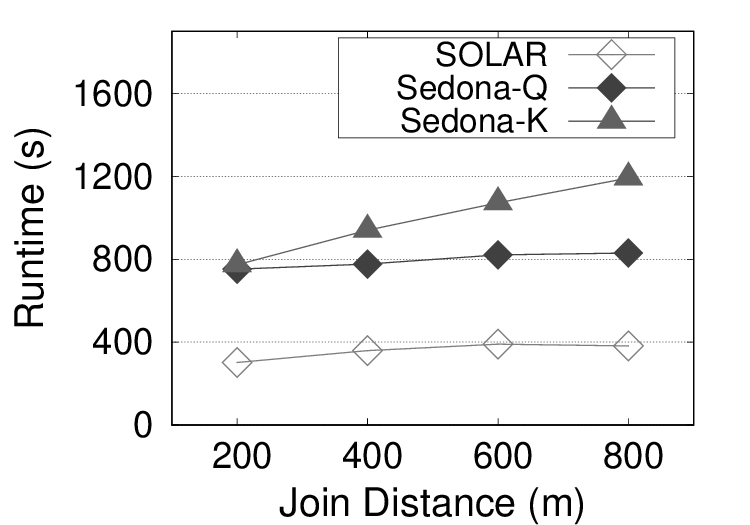}
    \caption{Worst Case}
    \label{var-base-worst-case-train}
  \end{subfigure}
  \vspace{-3mm}
  \caption{Runtime under Different Join Distances for \textit{training joins}}
  \vspace{-2mm}
  \label{trainvardist}
 
\end{figure*}

\begin{figure*}
  \begin{subfigure}[b]{.41\columnwidth}
    \includegraphics[width=\textwidth]{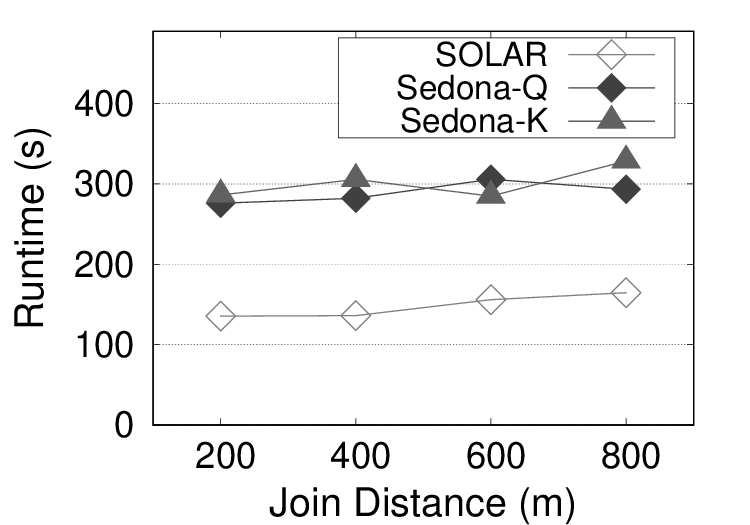}
    \caption{Best Case}
  \end{subfigure}
  \hfill
  \begin{subfigure}[b]{.41\columnwidth}
    \includegraphics[width=\textwidth]{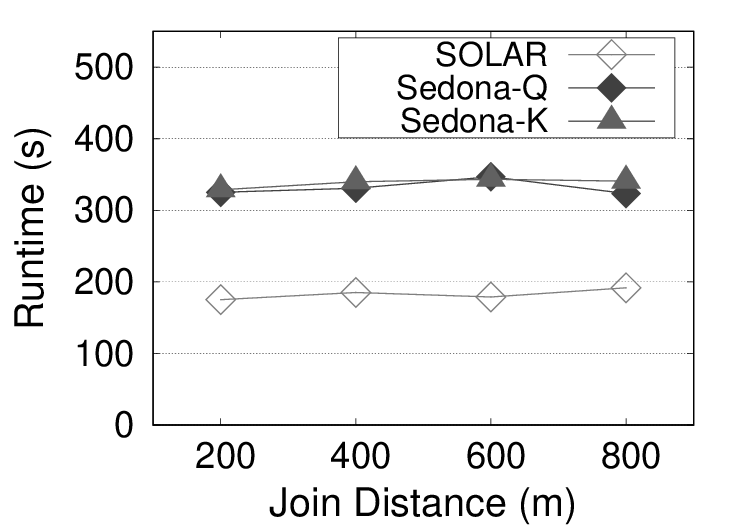}
    \caption{25 Percentile}
  \end{subfigure}
  \hfill
  \begin{subfigure}[b]{.41\columnwidth}
    \includegraphics[width=\textwidth]{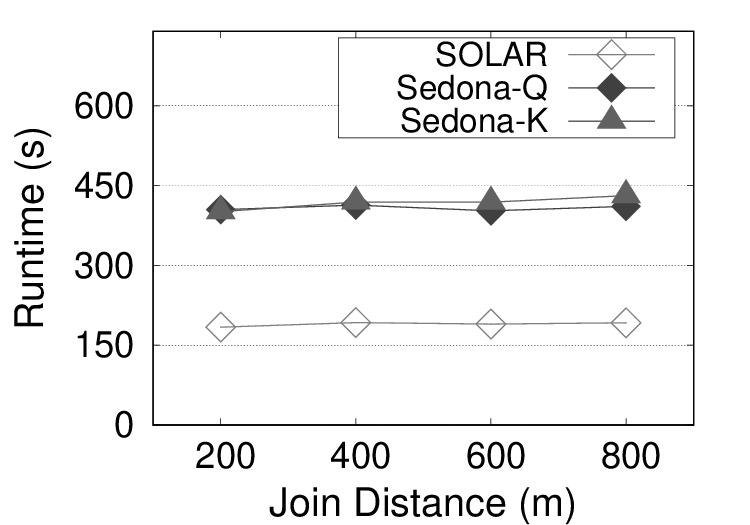}
    \caption{50 Percentile}
  \end{subfigure}
  \hfill
  \begin{subfigure}[b]{.41\columnwidth}
    \includegraphics[width=\textwidth]{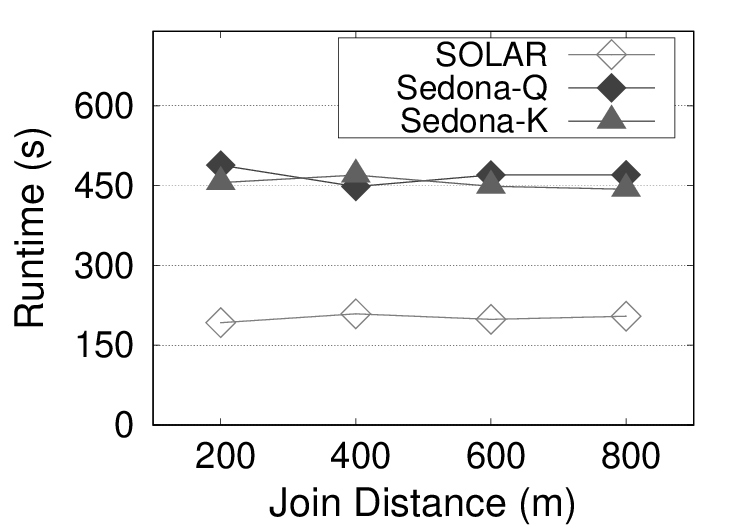}
    \caption{75 Percentile}
  \end{subfigure}
  \hfill
  \begin{subfigure}[b]{.41\columnwidth}
    \includegraphics[width=\textwidth]{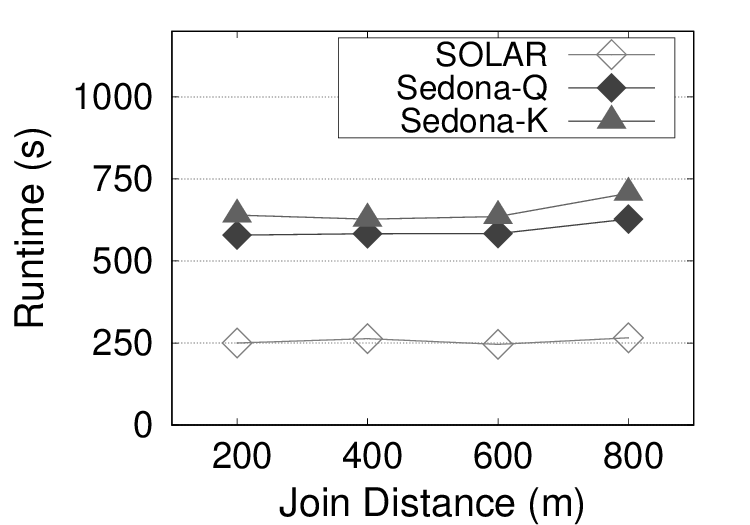}
    \caption{Worst Case}
    \label{var-base-worst-case-test}
  \end{subfigure}
  \vspace{-3mm}
  \caption{Runtime under Different Join Distances for \textit{test joins}}
  \vspace{-2mm}
  \label{testvardist}

\end{figure*}

\subsubsection{\textbf{Overhead in Partitioner Matching}}
When executing a new join, \textit{SOLAR} incurs some overhead due to the use of its learned model. This overhead involves using the Siamese Neural Network to match incoming datasets with existing partitioners, and then, based on the similarity score, using the decision-maker to determine whether an existing partitioner should be reused or if a new partitioner should be computed. Our experiments show that the minimum, median, and maximum overhead incurred during the partitioner matching step are 4.12ms, 5.25ms, and 14.29ms, respectively. The minimum, median, and maximum overhead incurred during the decision-making step are 10.84ms, 12.94ms, and 51.73ms, respectively. These minimal overhead times demonstrate that the computational cost associated with the partitioner matching procedure is negligible compared to the substantial runtime reductions.

\subsubsection{\textbf{Runtime Under Different Training Data Sizes}}
Figure~\ref{vary-training-percentage-train} and Figure~\ref{vary-training-percentage-test} illustrate the runtime performance of \textit{SOLAR} compared to the baseline algorithms when executing distributed spatial joins with varying percentages of training data. Specifically, we adjust the proportion of \textit{training datasets} utilized during the offline training phase to be 20\%, 40\%, 60\%, and 80\% of the total available datasets. As explained in Section~\ref{exp-queries}, the number of \textit{training datasets} directly affects the number of \textit{training joins} processed during the offline phase. Figure~\ref{vary-training-percentage-train} demonstrates that \textit{SOLAR} consistently outperforms the baseline algorithms in all cases for \textit{training joins}, achieving a maximum speedup that is up to 3.6X.  The speedup achievable by \textit{SOLAR} stems from its Siamese Neural Network's ability to detect similarity in the embeddings of training and test datasets. As we increase the size of the training datasets, the likelihood of \textit{SOLAR} detecting similarity to a previously seen partitioner for a spatially similar dataset increases. This allows for more partition reuse and greater computational savings. This capability eliminates the overhead associated with scanning data and performing partitioning operations at runtime.

Figure~\ref{vary-training-percentage-test} further shows that \textit{SOLAR} achieves a maximum speedup of 2.97X compared to the baselines for \textit{test joins}. This improvement results from the effectiveness of \textit{SOLAR}'s learning-based partitioner matching approach. Although the matched partitioner does not perfectly align with the distribution of new datasets, it exhibits high similarity. The time saved by avoiding on-the-fly partitioning significantly outweighs the additional overhead incurred by minor misalignments during local join computations, resulting in an overall runtime reduction. Furthermore, increasing the number of \textit{training datasets} exposes the Siamese Neural Network to a wider variety of dataset pairs. Consequently, the network becomes better at recognizing dataset similarities. Additionally, processing more joins during the offline phase results in storing more reusable partitioners. This increases the likelihood of successfully matching incoming joins to previously computed partitioners during the online phase. Collectively, the superior performance across both \textit{training joins} and \textit{test joins} highlights the effectiveness and practicality of \textit{SOLAR} in real-world settings.

\subsubsection{\textbf{Runtime Under Different Join Predicates}}
Figure~\ref{trainvardist} and Figure~\ref{testvardist} compare the runtime performance of \textit{SOLAR} and the baselines for workloads from \textit{training joins}, i.e., repeated joins and \textit{test joins}, i.e., unseen joins, under different join predicates, 
i.e., join distances. For \textit{training joins}, \textit{SOLAR} achieves runtimes up to 3.52X faster compared to the most efficient baseline. This speedup is attributed to \textit{SOLAR}'s ability to reuse previously encountered partitioners stored in the local repository. When a dataset in a join has been seen before, the model accurately matches the dataset to its corresponding partitioner, bypassing the need for real-time partitioning. For \textit{test joins}, \textit{SOLAR} achieves runtimes up to 2.69X faster compared to the best-performing baseline algorithm. This performance improvement demonstrates the model's strong generalization capability. Additionally, the speedup percentage achieved by \textit{SOLAR} is higher when the join distance is smaller. This is because partitioning overhead dominates the total runtime at smaller join distances, making the efficient reuse of existing partitioners particularly beneficial. Conversely, as the join distance increases, the local join operations become more computationally intensive, diluting the relative impact of partitioner reuse. Nevertheless, under practical spatial join predicates (e.g., within 1000 meters), the runtime improvements provided by \textit{SOLAR} remain substantial.

\section{Conclusion}
\label{conclusion}
This paper proposes \textit{SOLAR}, a scalable learning-driven approach to optimize distributed spatial joins. At its core, \textit{SOLAR} leverages a Siamese Neural Network to learn similarity representations from dataset embeddings. These learned representations enable the efficient reuse of effective partitioners when faced with new join queries. Our experimental findings consistently show that \textit{SOLAR} achieves significant speedups over baseline methods for both familiar and unseen join scenarios. For future work, we plan to accommodate other types of spatial joins including multi-way and kNN spatial joins. Beyond spatial joins, \textit{SOLAR}'s approach exemplifies a growing trend toward reuse-oriented and self-improving database systems. By embedding and comparing queries or datasets, such systems can selectively leverage prior work, including partitioning layouts, indexes, and even complete query plans. 

\bibliographystyle{ACM-Reference-Format}
\bibliography{sample}

\end{document}